\documentclass[12pt, a4paper]{article}

\usepackage{epsfig}
\usepackage{caption}
\usepackage{subcaption}
\usepackage[numbers, compress]{natbib}
\usepackage{graphicx}
\usepackage{rotating}
\usepackage{dirtytalk}
\usepackage{amscd}
\usepackage{amssymb}
\usepackage{amsbsy}
\usepackage{amsthm}
\usepackage[table]{xcolor}
\usepackage[toc,page]{appendix}
\usepackage{hyperref}
\usepackage{adjustbox}
\usepackage{titling}
\usepackage{authblk}
\usepackage{orcidlink}
\usepackage{float} 

\usepackage{chngcntr}
\counterwithin{figure}{section}
\counterwithin{table}{section}

\usepackage{vmargin}
\setmargins{3.0cm}{2.5cm}{15.5 cm}{22cm}{0.5cm}{0cm}{1cm}{1cm}

\theoremstyle{plain}

\theoremstyle{definition}

\theoremstyle{remark}

\providecommand{\keywords}[1]
{
  \small	
  \textbf{\textit{Keywords---}} #1
}

\newcommand{\appendixnumberline}[1]{Appendix\space}

\let\oldappendix\appendix
\makeatletter
\renewcommand{\appendix}{%
  \addtocontents{toc}{\let\protect\numberline\protect\appendixnumberline}%
  \renewcommand{\@seccntformat}[1]{Appendix~\csname the##1\endcsname\quad}%
  \oldappendix
}
\makeatother

\title{\vspace{-2cm}\textbf{Finding polarised communities and tracking information diffusion on Twitter: The Irish Abortion Referendum}}

\author{Caroline B. Pena$^{1}$\orcidlink{0009-0007-2329-8638},
Pádraig MacCarron$^{1}$\orcidlink{0000-0002-5163-9264},
David J.P. O'Sullivan$^{1}$\orcidlink{0000-0002-4754-3614}}  

\affil{$^{1}$ Mathematics Application Consortium for Science and Industry (MACSI), University of Limerick}

\date{caroline.pena@ul.ie}

\begin{document}


\begin{titlingpage}
    \maketitle
    \begin{abstract}
        The analysis of social networks enables the understanding of social interactions, polarisation of ideas, and the spread of information and therefore plays an important role in society. We use Twitter data --- as it is a popular venue for the expression of opinion and dissemination of information --- to identify opposing sides of a debate and, importantly, to observe how information spreads between these groups in our current polarised climate. 

        To achieve this, we collected over 688,000 Tweets from the Irish Abortion Referendum of 2018 to build a conversation network from users’ mentions with sentiment-based homophily. From this network, community detection methods allow us to isolate yes- or no-aligned supporters with high accuracy (90.9\%). We supplement this by tracking how information cascades spread via over 31,000 retweet-cascades. We found that very little information spread between polarised communities. This provides a valuable methodology for extracting and studying information diffusion on large networks by isolating ideologically polarised groups and exploring the propagation of information within and between these groups.
    \end{abstract} \hspace{10pt}

    \keywords{Social network, Twitter, information diffusion, information cascade, polarisation, community detection}
    
\end{titlingpage}

\newpage

\section{Introduction} \label{chap:introduction}

Twitter --- currently known as X\footnote{By the time of this analysis, X was still known as Twitter, therefore we will be referring to it as Twitter in this paper.} --- has facilitated public debate about a variety of subjects, and as a result, it has received considerable attention from researchers who wish to gain insights into the relationships and mechanisms that govern social interactions~\cite{cihon2016biased}. As of August 2023, Twitter had 450 million monthly active users, making it the $14^\textrm{th}$ most popular social media platform in the world in terms of users~\cite{shewaleTwitterStatistics2023}. In Ireland, by 2022,  1.35 million people --- 32.6\% of the over-13 year-old audience --- used Twitter~\cite{kempTwitterIreland2022}.

Twitter is a popular venue for political discussions~\cite{bestvater2022politics, flamino2023political}, such as referendums~\cite{grvcar2017stance, del2017news, furman2020end, o2017integrating}. Ireland has a reasonably lengthy history of dealing with referendums~\cite{barrett2020some}, and two of them --- the Irish same-sex marriage and the Irish abortion referendums --- have received a lot of attention in social media, where millions of tweets were shared, and many public figures expressed their opinion~\cite{murphy2016marriage, hunt2019twitter}.  

Previous research by O'Sullivan et al.~\cite{o2017integrating} examined the $2015$ Irish Marriage Referendum and successfully used Twitter data and social network analysis to identify groups of users who were pro- and anti-same-sex marriage equality with a high degree of accuracy. Our research improves upon this work in two ways. Firstly, we demonstrate that we can obtain better classification accuracy of users into polarised communities on two independent datasets (the Irish Marriage referendum of 2015 and the Abortion referendum of 2018) while using substantially less data, which is crucial given the cost of data gathering. Secondly, we extend the previous analysis by tracking not only how yes- and no-supporters of the referendum interact individually but how the information they share spreads across the network, within and between communities via the construction of retweet cascades~\cite{goel2016structural, vosoughi_spread_2018, juul_comparing_2021}, that is, a structure to describe who shared the same content from whom, following a timeline and relationships among users. 

In summary, we address the following questions in this work:

\begin{enumerate}
  \item Is it possible to extend O'Sullivan et al.'s~\cite{o2017integrating} analysis using different methods and less data (excluding the followership data)?
  \item How does information spread within and between ideological communities?
\end{enumerate}

Key to our analysis is to show that users tend to group together according to the language they use in the discussion. The tendency that nodes have to associate preferentially with nodes which are similar to themselves in some way is a common phenomenon in many social networks, as in the classic school example, where children tend to form groups based on characteristics in common, like age~\cite{newman2003structure}. This kind of selective linking is called \textit{assortative mixing} or \textit{homophily} in social networks. In the context of a referendum, where people either vote in favour or against it, we expect a polarised environment --- a self-reinforcing system where users tend to talk to others that share the same opinions much more frequently than to whom share opposite ideas~\cite{smith2024polarization} --- and the presence of assortative mixing together with ideological label assignments is a good proxy for polarisation~\cite{taylor2018exploring}. In this sense, we will use a combination of methods to uncover assortative mixing and assign labels to the groups.

A combination of social-network structure and sentiment analyses have been used as a tool to understand and even predict human behaviour around many topics~\cite{thelwall2017heart}, such as stock market fluctuations~\cite{ranco2015effects, zheludev2014can}, tracking the spread of viruses~\cite{alamoodi2020sentiment, crable2020exploring, lim2017unsupervised, raamkumar2020measuring, singh2018sentiment}, understanding the results of elections~\cite{bermingham2011using, budiharto2018prediction, ramteke2016election}, and are powerful tools to uncover homophily~\cite{caetano2018using, de2011tie, yuan2014exploiting}. In this work, we combine the analyses of sentiment and social structure to explore Twitter conversations about the Irish marriage referendum and the Irish abortion referendum.  

In O'Sullivan et al.~\cite{o2017integrating}, a sentiment analysis tool~\cite{thelwall2012sentiment} was used to quantify the positive and/or negative emotions of each tweet. This information was later aggregated to calculate the sentiment of each user in the network. The sentiment information was then used in conjunction with a combination of the conversation (who mentions whom in a tweet) and the followers (who are friends with whom on Twitter) networks to uncover community structure, which showed that the Irish Marriage referendum was polarised with a large echo-chamber effect --- with a small group of users who communicated across community boundaries. This analysis relied heavily on the gathering of followership data, which is a time and fund consuming task. Therefore, our first aim in this work is to show that it is possible to use only the conversation graph --- and not a combination of this and the followership graph --- and sentiment to find communities of yes- and no-supporters.

Once we have these groups, we extend previous work and give an example of the utility of the classification method, analysing the behaviour of cascade trees to understand how the community structure  influences the spread of information in the social context. Cascade trees allow us to identify where the spread originated and to examine the structure it created. The link between community structure and how information spreads through a network has been reported to be a major factor in the spreading of content~\cite{weng2013virality}. The spread of behaviour and information in social networks has been largely studied~\cite{centola2010spread, keating2022multitype, gleeson2020branching}, as well as on a wide variety of social networks such as Facebook~\cite{dow2013anatomy, sun2009gesundheit}, Reddit~\cite{deza2015understanding}, Google+~\cite{guerini2013exploring}, TikTok~\cite{alonso2021beyond}, and LinkedIn~\cite{anderson2015global}, to name a few. We are particularly interested in how information spreads within and between communities in polarised conversation networks on Twitter.

To examine the online referendum's social network structure, its interdependence with sentiment and the side in the debate that a user supports, we analyse two different data sets. One containing tracked tweets associated with the Repeal the $8^\textrm{th}$ Referendum --- or RT8 --- (our focus in this paper)  and the second containing tracked tweets associated with the Irish Marriage Referendum --- referred to as IMR (presented in Appendix \ref{chap:IMR}). The Irish Marriage Referendum dataset is the same one utilised by O'Sullivan et al.~\cite{o2017integrating}.

In the remainder of the main text we will explain our methodological approach (Section~\ref{chap:methodology}), followed by our case study on the Irish Abortion Referendum, where we show how we gathered the data and conduct a brief statistical analysis on it (Section~\ref{chap:data}), show how the network structure was created and how we incorporated sentiment scores into the network (Section~\ref{chap:mention_net}), analyse the correlation between sentiment scores and the social structure (Section~\ref{chap:sentiment}), which suggests the existence of polarised communities (Section~\ref{chap:communities}). Using the polarised groups, we show our findings on the diffusion of information inside and between ideological communities (Section~\ref{chap:cascades}). The main part of this work ends with the conclusions around the Repeal the $8^\textrm{th}$ referendum discussion and an outline of limitations and future work (Section~\ref{chap:conclusion}).

\section{Methodology} \label{chap:methodology}

In this section we provide a step-by-step guide for how the analysis will take place in the subsequent case-study section. We start by collecting data. The type of data of interest is data that has a natural division of the individuals. Conversation data around controversial topics --- for example, referendums or elections --- are of this type. We gather tweets about the Irish Abortion Referendum of 2018 and the Irish Marriage Referendum of 2015. The data must encompass different perspectives of the debate; therefore the hashtags must be carefully selected, i.e., hashtags that are commonly used and informative of (and unique to, if possible) the matter of interest, at the same time that they are representative of the different opinions from different sides of the conversation. As the focus is on the conversation that takes place between users, a conversation network must be built, where the nodes are the Twitter users in the collected dataset, and the links (edges) are mentions from one Twitter user to another regarding the matter being researched, where a mention represents one user either trying to engage another user in a dialogue, draw their attention to a piece of content (URL for example), or a reply to their tweet. 

It is worth noting that most users will have a small number of tweets in the network, as they rarely interact with other users. The analysis must be made on the group of users who are highly active in the online discourse; therefore, we filter the network by only including a user if they have at least one mutual mention with other user(s). This results in a set of users who are more active than typical users on the topic and are more representative of those who engage with other users. This filtering has the added benefit of leaving us with a set of users who have more tweets on average to analyse the language that they use to express sentiment about the topic. Secondly, we maintain only the nodes that are in the largest strongly connected component of the network, which ensures that every node is connected to every other node, i.e., it is possible to reach any node in the network by starting from any random node.

\begin{figure}[h] 
    \centering
    \includegraphics[width=0.95\textwidth]{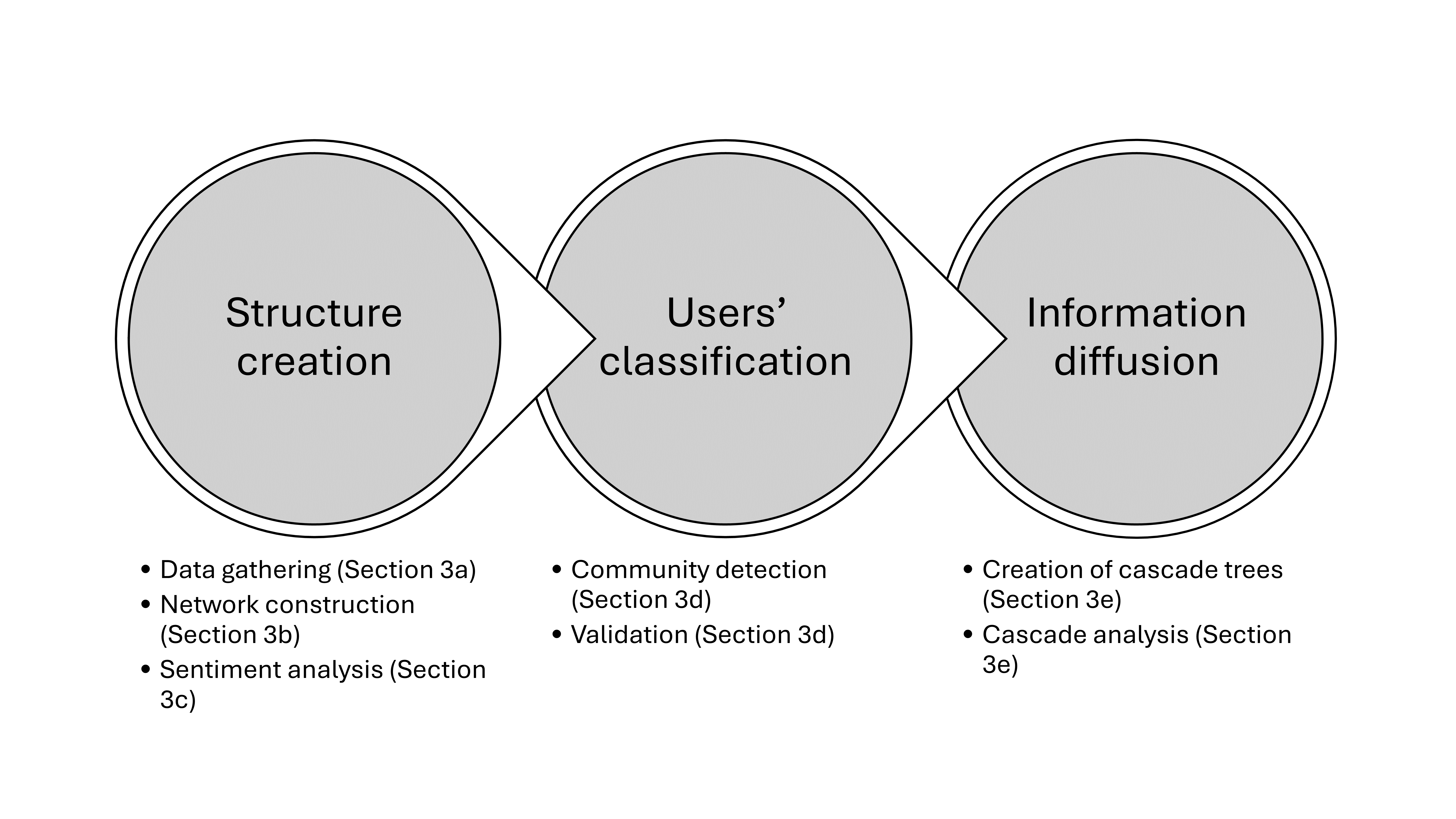}
    \caption{Schematic of our methodological approach.}
    \label{fig:agenda}
\end{figure}

A user's sentiment represents how positive or negative their language is. Sentiment may be used as a proxy for homophily, i.e., users tend to group together according to their average sentiment (positive or negative). Therefore, one must first extract the sentiment from the collected tweets. We do this by using natural language processing to assign scores to each text content sent between pairs of users. We use a lexicon developed for short text that assigns negative scores to words that suggest negative language, and positive scores to words that indicate positive language. The sentiment scores may vary considerably among tweets sent by a user, making this measurement noisy. Therefore, to extract a user sentiment, we calculate the average sentiment of all tweets sent by the user.

Randomisation tests may show sentiment alignment among users. If the results are not due to chance, sentiment can be used as a proxy for homophily. We analyse the correlation between the sentiment that a user sends to others and the sentiment they receive from others, and test if users tend to connect to (mention) others that send the same type of sentiment (positive or negative) in the network. In this setting, it is reasonable to assume that there is a correlation between yes and no supporters with positive and negative sentiment, respectively, as they will naturally use more positive or negative language to express their opinions about the referendum, i.e., a yes supporter may use phrases like ``I support this'', ``I agree with it'', ``this is great'', while a no supporter may say things like ``I disagree with the proposal'' , ``this is not right''. However, in more complex settings where this might not be as clear, the network structure of who tends to talk to whom will help uncover groups of people that share similar opinions.

Therefore, to uncover the polarisation structure, standard methods on graph theory that allow the identification of groups of nodes can be used. Community detection algorithms analyse the network topology and detect clustering patterns --- groups of users that are highly connected among themselves and scarcely connected to nodes in another group --- assigning each node a membership to a group of nodes. The community detection algorithms used in this work --- fast greedy, Louvain, Leiden, spinglass, walktrap, leading eigenvector, infomap, and label propagation --- allow link weights to be used for the community detection. The link weights may increase the accuracy of the algorithm. Because sentiment alignment can be used as a proxy for homophily, as seen by previous randomisation tests, and because nodes in the same group do tend to share similar characteristics, i.e., sentiment homophily, we assign the users' average sentiment as link weights. However, weights cannot be negative. Therefore, we use the absolute value of users’ sentiment to determine the link weights in the network. This approach is consistent because users who express very negative sentiments are positioned far from those who express very positive sentiments in the network topology. We then choose the algorithm that performs the best based on modularity. The user classification is validated by annotated data, where we select a sample of the users. For annotation, the researcher is blinded to the user name and community but classifies them as a yes/no supporter based on all the tweets they sent to others regarding the studied referendum. Following this, error metrics can be calculated, such as balanced accuracy, to assess the method's performance.

With the groups detected, one can then analyse the interactions of the users in light of the group labels. For example, we analyse how information spread between and inside these groups. We do so by extracting the information diffusion cascades by using text and time information and the mentions network topology (the full dataset on mentions, without any filtering to avoid biases). We are then able to study who shares the same content from whom and add information on the membership from the community detection step to analyse how information spreads within and between polarised communities.

In Section 3 we apply our method on the Twitter data for the Irish Abortion Referendum of 2018, explaining in detail the theory behind each step taken. Figure~\ref{fig:agenda} illustrates our methodological approach.

\section{Case study: the Irish Abortion Referendum of 2018}\label{chap:case_study}

\subsection{Background and data description} \label{chap:data}

The referendum concerning abortion legislation, the so-called Repeal the 8$^\textrm{th}$ Referendum, was one of the most recent referendums held in Ireland, on 25 May 2018. It abolished the 8$^\textrm{th}$ Amendment of the Constitution --- which gave the unborn equal rights to the mother and prohibited abortion in the State --- and replaced it with the following wording~\cite{lapowskyEvWilliamsTwitter2013}:

\vspace{1cm}

\hfill\begin{minipage}{\dimexpr\textwidth-2cm}
\say{Provision may be made by law for the regulation of termination of pregnancies.}
\end{minipage}

\vspace{1cm}

The Repeal the 8$^\textrm{th}$ Referendum was one in a series of referendums about the matter held in Ireland since the inclusion of the 8$^\textrm{th}$ Amendment in 1982~\cite{field2018abortion}. In 1992, two amendments passed by referendums, which overturned the decisions preventing the right to travel for abortion and the right to distribute information about abortion services in other countries, and one amendment proposal --- to exclude the risk of death by suicide as grounds for abortion --- was rejected. In 2002, there was another attempt to exclude risk of death by suicide as grounds for abortion, and also to create explicit criminal penalties for abortion. This bill did not pass either. The abortion discussion remained outside the mainstream of Irish politics until 2012, when Savita Halappanavar, an Indian dentist living in Ireland, had the termination of her pregnancy denied in an Irish hospital and died from complications arising from a septic miscarriage~\cite{field2018abortion, bbcSavita}. Protests and public discussion~\cite{bbcSavita, irishtimesSavita} culminated in the Repeal the 8$^\textrm{th}$ Referendum in 2018. The turnout of this referendum was $64.13\%$, making it the fourth-highest turnout for any referendum in the history of the State~\cite{field2018abortion}. The ``Yes'' side won by a margin of $66.4\%$ to $33.6\%$~\cite{bbcRepeal}. Also, with $1\,429\,981$ votes in favour, it received the biggest popular endorsement of any referendum question on abortion to date in the Irish State~\cite{field2018abortion}.

The tweet data on the Repeal the $8^\textrm{th}$ Referendum~\cite{Pena_Twitter_data_on_2024} was collected using the Twitter Academic Research API, which returns precise, complete, and unbiased data from the Twitter archive~\cite{TwitterAPI}. However, as the data was collected a few years after the referendum, there could be missing tweets due to deletion or removal of the user's account.

\begin{figure}[h]
    \centering
    \includegraphics[width=0.9\textwidth]{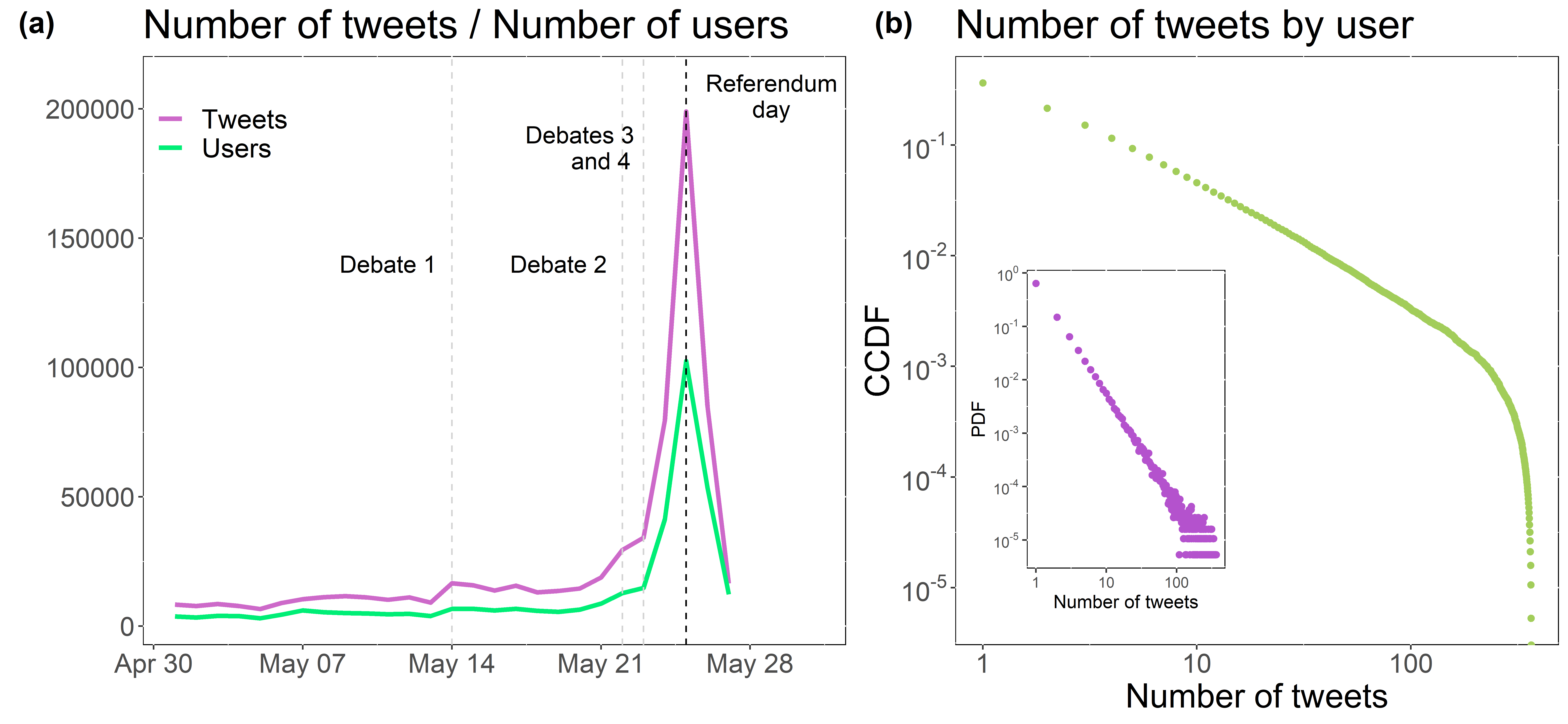}
    \caption{Data summary for the Repeal the $8^\textrm{th}$ discussion. (a) Repeal the $8^\textrm{th}$ activity time frame showing the televised debates and the day of the referendum. Note a high peak on the referendum day. (b) Complementary cumulative distribution function for the number of tweets per user and an inset of the probability distribution function of the same data.}
    \label{fig:RT8_data_summ}
\end{figure}

For the Repeal the $8^\textrm{th}$ Referendum, every tweet containing the hashtags \#repealthe8th, \#savethe8th, \#loveboth, \#together4yes, and \#retainthe8th from the $1^\textrm{st}$ of May to the $27^\textrm{th}$ of May 2018 (two days after the referendum) was collected. Those hashtags in particular were chosen given their representativeness of both sides of the discussion and their popularity (see Appendix~\ref{chap:hashtags} for more information on how we selected the hashtags). Four debates around the Repeal the $8^\textrm{th}$ discussion were televised, during which small increases in tweet numbers were recorded, and these can be seen in Figure~\ref{fig:RT8_data_summ}~(a). A large peak in the days leading up to the referendum day is also observed. The data collected contains the text of the tweet, the date and time when the tweet was created, whether the tweet was a retweet or a reply (or neither, i.e., original), hashtags and mentions contained in the text. This information will be used to build the mentions network, which will be explained in Section~\ref{chap:mention_net}, and to retrieve retweet cascades (Section~\ref{chap:cascades}).

Figure~\ref{fig:RT8_data_summ}~(b) shows that the number of tweets per user has a heavy tailed distribution. The vast majority of users only posted a small number of tweets with the tracked hashtags, while a small number of users were responsible for a large volume of tweets. Out of the gathered tweets, $13.96\%$ of the tweets were original posts, retweets represented $79.57\%$ and replies represented $6.47\%$ of the tweets (Table~\ref{table:summ_data_description}). This shows that there was strong engagement among users in our dataset --- the majority of tweets were retweets and replies --- despite the heavy tail. The large proportion of retweets is indicative of the use of Twitter for information spread, which will be the focus of Section~\ref{chap:cascades}.

\begin{table}[h!]
\centering
\caption{Summary table of gathered dataset (RT8).}
\begin{tabular}{ll}
\hline
Attribute & Value\\
\hline
Period analysed & 1-27 May 2018\\
No. users & 188,928\\
No. tweets & 688,903\\
No. original tweets & 96,182\\
No. retweets & 548,147\\
No. replies & 44,574\\
\end{tabular}
\label{table:summ_data_description}
\end{table}

In order to analyse user behaviour around the topics in question, we need to understand how users connect and build the conversation network structure. The next section will explain how the network structure was created.

\subsection{Network structure creation} \label{chap:mention_net}

Our goal is to construct a network that is reflective of the conversations taking place on Twitter around the Repeal the 8$^\textrm{th}$ Referendum. For a conversation to take place, we require a network based on the mentions, that is, who talks to whom about the referendum. On Twitter, a mention to a user, text starting with the ``@'' sign, can be generated by tagging the user in a tweet or by replying to the user (note: quoting tweets do not create explicit mentions as they do not create a link to the user that originally tweeted the content via an ``@'' sign). We are interested in every form of mentioning among users, therefore we draw links connecting users by any type of mention (either tags or replies). It is also important to clarify that the mentions network is the observed communication between users rather than a structure underlying the possibility of connections. This is important in the next steps as we want to capture how users are connected and active in the network in order to detect polarised communities.

In this step we keep $177\,324$ unique users --- $11\,604$ users did not mention anyone in our gathered tweets. Once we have the users (nodes) connected by who mentions whom, we want to filter the users that 1) are more active and engage with the subject~\cite{charlton2016mood, grindrod2016comparison, beguerisse2017and, o2017integrating}, and will have more tweets to average the sentiment over, 2) have strong social connections, which can be used to uncover community structures --- politically active users tend to organize into homogeneous communities~\cite{conover2011political}, which might be indicative of polarisation. Therefore, we maintain only the nodes that have at least one mutual mention (i.e., A mentions B and B mentions A), creating the mutual mentions network (the network was downsized to $4\,967$ nodes at this point). It is important to stress here that users do not necessarily have mutual mentions with all users they talk to; they only need mutual mention(s) with at least one other user in the network. We also filter the mutual mentions network into its largest strongly connected component, so that we have a network with active users that are strongly connected. The final mutual mentions network filtered by its largest strongly connected component has $3\,977$ nodes. It is important to highlight here that real data is known to be complex and noisy, therefore applying filters is a common practice in data analysis. Our aim is not to analyse the entire dataset originally scraped, but to build a concise network with only the most active and interconnected users to ensure that useful and clear results are generated. Without the filtering steps, the sentiment analysis would be compromised as many users in the network would have sent but not received tweet(s) from others, making it impossible to evaluate the correlation between their sentiment-in and sentiment-out. Furthermore, our ability to find polarised communities would be highly affected due to several nodes being peripheral in the discussion, users which do not present a clear behaviour that allows them to be classified into one of the polarised groups. This is both due to their weak connection to the core and high active group, and the small number of tweets they send to others, making it difficult to draw a conclusion about their political positioning in the debate.

\begin{table}
    \centering
    \caption{Summary statistics for the full and the mutual mentions networks (RT8).}
    \begin{tabular}{lrrrrr}
    \hline
    Network & Nodes & Links & Reciprocal links & Avg. out-degree & Transitivity\\
    \hline
    Full & 8\,154 & 217\,136 & 44\,155 & 26.63 & 0.07\\
    Mutual & 3\,977 & 162\,416 & 44\,002 & 40.84 & 0.12\\
    \end{tabular}
    \label{tab:full_mutual}
\end{table}

Table~\ref{tab:full_mutual} compares the largest strongly connected component of the full mentions network (the mentions network before filtering by the nodes with reciprocal mentions) with the largest strongly connected component of the mutual mentions network. It shows that the mutual mentions network contains users that have more connections (higher average out-degree) and are highly connected to each other (higher transitivity). Here, transitivity means that if a user mentions two other users, they also mention each other. This is an important feature when applying community detection methods to find polarised communities. From here on, when we refer to the mentions network, we are referring to the largest strongly connected component of the mutual mentions network.

\begin{figure}[h] 
    \centering
    \includegraphics[width=0.9\textwidth]{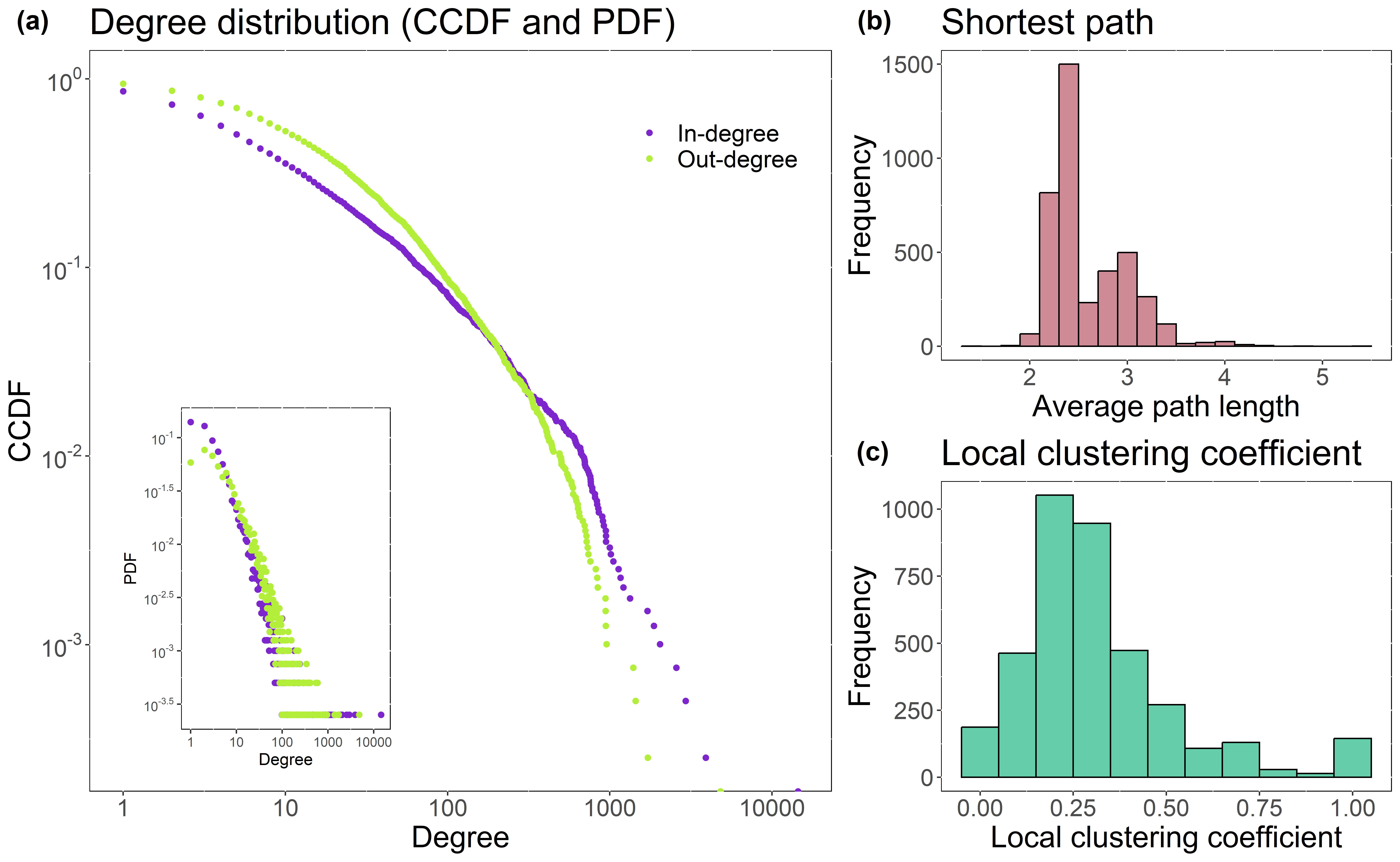}
    \caption{(a) Complementary cumulative distribution function (CCDF) for in-degree (purple) and out-degree (green) per user in the mutual network, and inset of the probability distribution function (PDF) of the same network; (b) Average shortest path; (c) Local clustering coefficients.}
    \label{fig:RT8_stats}
\end{figure}

Figure~\ref{fig:RT8_stats}~(a) shows that, as expected for social networks~\cite{newman2003social}, our mentions network has a heavy-tailed degree distribution. The curves show the complementary cumulative distribution function of the in-degree and the out-degree by user, with the inset showing the probability distribution function of our mentions network. A high number of users send (out-degree, green) a small number of tweets, while a small number of users post (and mention) a large volume of tweets. The same is observed for the quantity of tweets received (in-degree, purple) by each user. 

The local clustering coefficient (Figure~\ref{fig:RT8_stats}~(c)), suggests that the users are strongly interconnected. Almost 20\% of the users have a local clustering coefficient greater than 0.5, or in other words, almost 20\% of the users have neighbours with a probability greater than 50\% of being connected together.

Figure~\ref{fig:RT8_stats}~(b) shows the mean shortest distance between vertex pairs. The spread of information is faster in networks that have a small average
path length~\cite{newman2003structure}: if it takes only a small number of steps for information to spread from one person to another, then the information will propagate much faster compared to a network with a large average number of steps between nodes. The most frequent average path length between nodes observed in our mutual network for the RT8 data is 2-3 steps, suggesting that the spread of information occurs quickly among users, which will be further explored in Section~\ref{chap:cascades}.

\subsubsection{Sentiment scores}\label{subchap:sentiment}

Emotions are important in communication, and large-scale studies of communications need methods to detect emotions~\cite{thelwall2017heart}. For this purpose, various sentiment lexicons have been developed in the past few years in an attempt to better characterise and treat sentiment from texts~\cite{agarwal2011sentiment, chmiel2011collective, liu2012sentiment, loughran2011liability, mohammad2013crowdsourcing, saini2019sentiment, thelwall2012sentiment}. Sentiment analysis has been used as a tool to understand and even predict human behaviour around many topics~\cite{thelwall2017heart}, such as financial markets~\cite{ranco2015effects, zheludev2014can}, tracking the spread of viruses~\cite{alamoodi2020sentiment, crable2020exploring, lim2017unsupervised, raamkumar2020measuring, singh2018sentiment}, film box-office performance~\cite{apala2013prediction, kim2018text, rajput2017box}, reviews~\cite{nagamma2015improved}, and even the analysis of elections~\cite{bermingham2011using, budiharto2018prediction, ramteke2016election} and storm impacts~\cite{spruce2020using}, and more recently the effects of COVID-19 outbreaks on mental health~\cite{barkur2020sentiment, valdez2020social}. Also, recent studies have successfully pointed out how sentiment can be used to help model the spread of information on Twitter~\cite{o2017integrating,chmiel2011collective, alvarez2015sentiment}.

In our study, sentiment is used to capture stance and help understand users' general feelings about the topic. It is used 1) as a means of understanding how users that share the same (opposite) type of sentiment connect in the network, and 2) in combination with the mentions network topology to find communities of yes- and no-supporters. While O'Sullivan et al.~\cite{o2017integrating} made use of \textit{SentiStrength}~\cite{thelwall2012sentiment} and its built-in lexicon for sentiment analysis, we chose to use the R package \textit{tidytext}~\cite{silgeTextMiningTidy2017} in this study due to its usability and easy implementation with large amounts of text. The lexicons in the \textit{tidytext} package are based on unigrams, i.e., single words, meaning the respective algorithms check word by word to assign scores. 

We use \textit{AFINN}~\cite{nielsen2011new}, a lexicon constructed specifically for microblogs, therefore appropriate to analyse tweets. Furthermore, \textit{AFINN} performs well in polarisation analyses on Twitter, according to Hern{\'a}ndez et al.~\cite{hernandez2014iradabe}. It assigns the words a score that ranges between $-5$ and $+5$, with negative scores indicating negative sentiment and positive scores indicating positive sentiment. The sum of positive and the sum of negative scores were then obtained for each tweet. Finally, the positive scores were re-scaled in order to fall into the interval from $0$ to $5$, and the same was done for the negative scores, with the range of scores to be between $-5$ and $0$, instead. The final sentiment score for a tweet is the sum of its negative and its positive scores. Tweets containing neither positive nor negative scores are assigned a sentiment score equals to zero (neutral).  Figure \ref{fig:sentiment} in Appendix~\ref{chap:appendix} exemplifies how the mentions network was built. On the network topology, each mention tweet between two users is represented as a link in the mentions network, and each one of those links now has an attribute containing the sentiment score of its text. In the next section we will use those sentiment scores and the network structure to understand how users communicate according to the sentiment they send and receive from others.

\subsection{User sentiment and social structure} \label{chap:sentiment}

Sentiment analysis models may struggle with a few characteristics of opinionated texts, such as 1) identifying sarcasm or irony, which are common in written language, 2) adequately handling negations or contrasting statements, and 3) managing domain-specific sentiments or jargon~\cite{serrano2015sentiment, pang2008opinion}. Therefore, the sentiment score obtained for a single tweet may provide unreliable information about the user's general expressed sentiment toward a topic.

To obtain a more robust indication of the users' sentiment, the scores of all the tweets produced by one user were aggregated to obtain a single score (average of the scores obtained from all tweets by each user in the mutual mentions network), as exemplified in Figure~\ref{fig:sentiment} in Appendix~\ref{chap:appendix}. Those users are highly active (given our process of constructing the mutual mentions network), therefore our users' sentiments are averaged over a number of tweets, making it a robust measurement. The rescaled scores obtained with the AFINN lexicon are used to calculate the average sentiment in- (of all tweets received) and out- (of all tweets sent) of each user. In Section~\ref{chap:cascades} these scores will form an important part of classifying users as Yes or No supporters.

\begin{figure}[t] 
    \centering
    \includegraphics[width=0.9\textwidth]{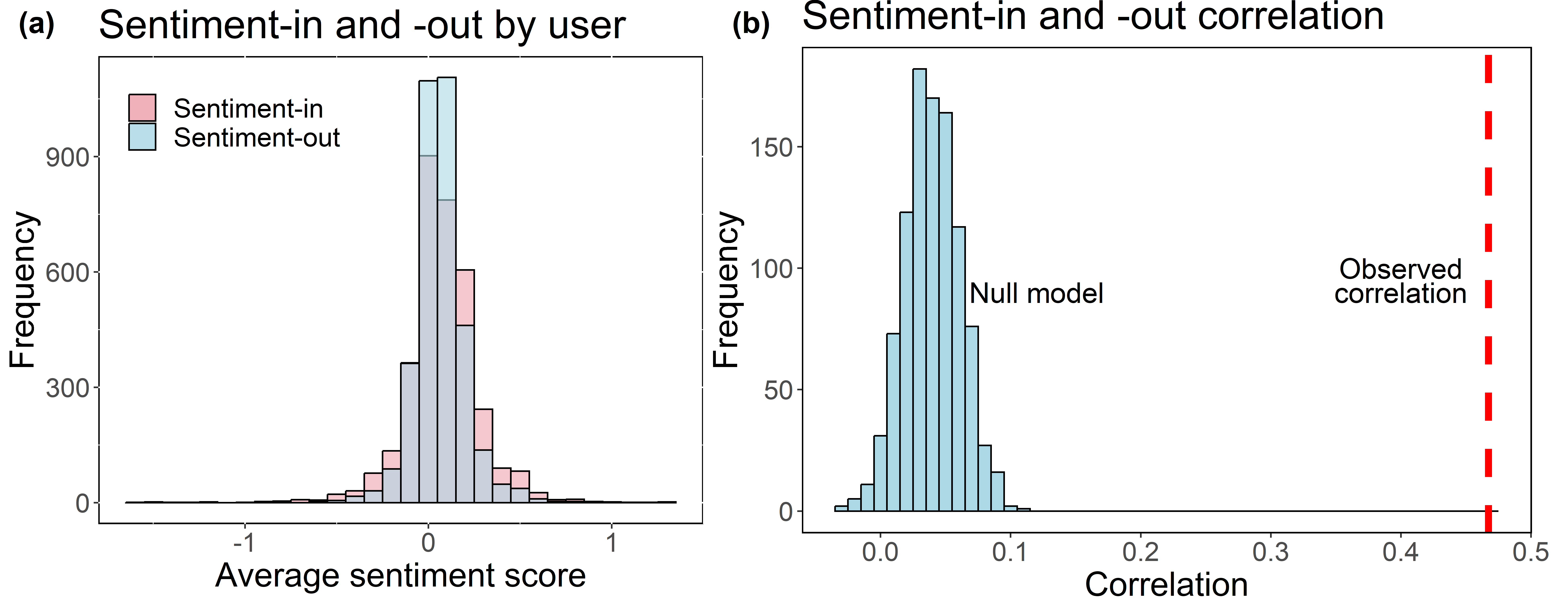}
    \caption{(a) Distribution of average sentiment-in by user in blue, and average sentiment-out by user in pink; (b) Distribution of simulated correlation values between the sentiment-in and sentiment-out by user, and the observed correlation. The blue bars are the resultant correlation distribution obtained after 1\,000 simulations, and the red dashed line represents the observed correlation.}
    \label{fig:sentiment_average}
\end{figure}

Both users' in-sentiment and out-sentiment scores distributions are centred on zero and slightly right skewed (Figure~\ref{fig:sentiment_average}~(a)). Regarding the in-sentiment scores distribution obtained by the RT8 data, $93.76\%$ of the users send tweets scoring between $-0.4$ and $+0.4$ on average, $71.79\%$ of which are between $-0.2$ and $+0.2$. The out-sentiment distribution records $96.91\%$ of the users scoring an average sentiment-out between $-0.4$ and $+0.4$, $83.93\%$ of which are between $-0.2$ and $0.2$ interval. Also, $65.75\%$ of the users record a positive in-sentiment average score, while $70.66\%$ of the users send positive tweets to others. 

These results raise two main questions: 

\begin{enumerate}
  \item Are user in- and out-sentiment scores correlated?
  \item Do users whose tweets have similar sentiment tend to be grouped in the network?
\end{enumerate}

If the sentiments of the tweets that a user sends and receives are correlated, and users tend to cluster together with others that share similar sentiment, we could then consider sentiment alignment as a proxy for homophily among users. We can reasonably expect this as users with a similar disposition towards the referendum may communicate using similar language~\cite{xia2024russian, o2017integrating}. On a user level, the randomisation tests analyse how strong sentiment is as a signal, i.e., if sentiment was very noisy, the correlation between the sentiment that a user sends and the sentiment they receive from neighbours would also be very noisy and therefore be a poor measure of similarity of connected users.

To answer (i), the correlation between users' in- and out-sentiment is examined. The observed Pearson correlation between sentiment-in and sentiment-out score distributions is $0.4672$, which indicates a moderate positive linear relationship between these two attributes ~\cite{benesty2009pearson, taylor1990interpretation}. To confirm that this correlation is not due to chance, a randomisation procedure based on redistributing the sentiment of a user's tweets is undertaken. The randomisation procedure is as follows:

\begin{enumerate}
\item Sample a sentiment score for each connection from the reciprocal network with replacement. Here, we use sampling with replacement to make sure the probability of selecting any particular sentiment score remains the same in future draws. This procedure also keeps the network topology intact.
\item Calculate the average randomised in- and out-sentiment of each user.
\item Calculate the correlation coefficient between the average randomised in- and out-sentiment distributions.
\item Repeat steps 1-3 1\,000 times to create the null model for where the correlation coefficient could lie if there was no connection between the in- and the out-sentiment for each user.
\end{enumerate}

Figure~\ref{fig:sentiment_average}~(b) shows the comparison of the resulting distribution of the correlation between the average randomised in- and out-sentiment scores after $1\,000$ iterations of the randomisation procedure with the observed correlation of sentiment-in and sentiment-out scores on the observed mutual mentions network for the Repeal the $8^\textrm{th}$ Referendum --- $0.47$. The fact that the correlation obtained from the randomisation is far away from the observed correlation indicates that there is a non-trivial correlation between the sentiment of what a user sends and receives. The observed correlation between sentiment-in and sentiment-out scores suggests that users may be more likely to be connected to other users with similar sentiment scores (meaning that they tend to use similar language about the referendum). On the network topology level, this is an indication that users may be grouped together according to the sentiment score, i.e., if the sentiment-in and -out of a user are correlated, this user is connected to others that send to them similar sentiment as to what they send out to others.

Therefore, to assess if users tweeting similar sentiment tend to be grouped in the network, three class labels are created for users according to their sentiment-aggregate scores: above zero are ``positive'', less than zero are ``negative'' and scores equal to zero are ``unknown''. We used these labels to find the fraction of connections between users of these classes. We denote the fraction of links between positive and positive users as \textit{fpp}, the fraction of links between positive and negative users as \textit{fpn}, between positive and unknown users as \textit{fpu}, and so on. In total, there are nine types of links: \textit{fuu}, \textit{fup}, \textit{fun}, \textit{fnn}, \textit{fnu}, \textit{fnp}, \textit{fpp}, \textit{fpu}, and \textit{fpn}. We then randomise the class labels of each user by sampling from the observed distributions with replacement, and recalculate the fraction of connections; we repeat this process $1\,000$ times. As before, we compare the randomised distributions of the fractions with the observed fraction in our data.

\begin{figure}[t] 
    \centering
    \includegraphics[width=8cm, height = 6cm]{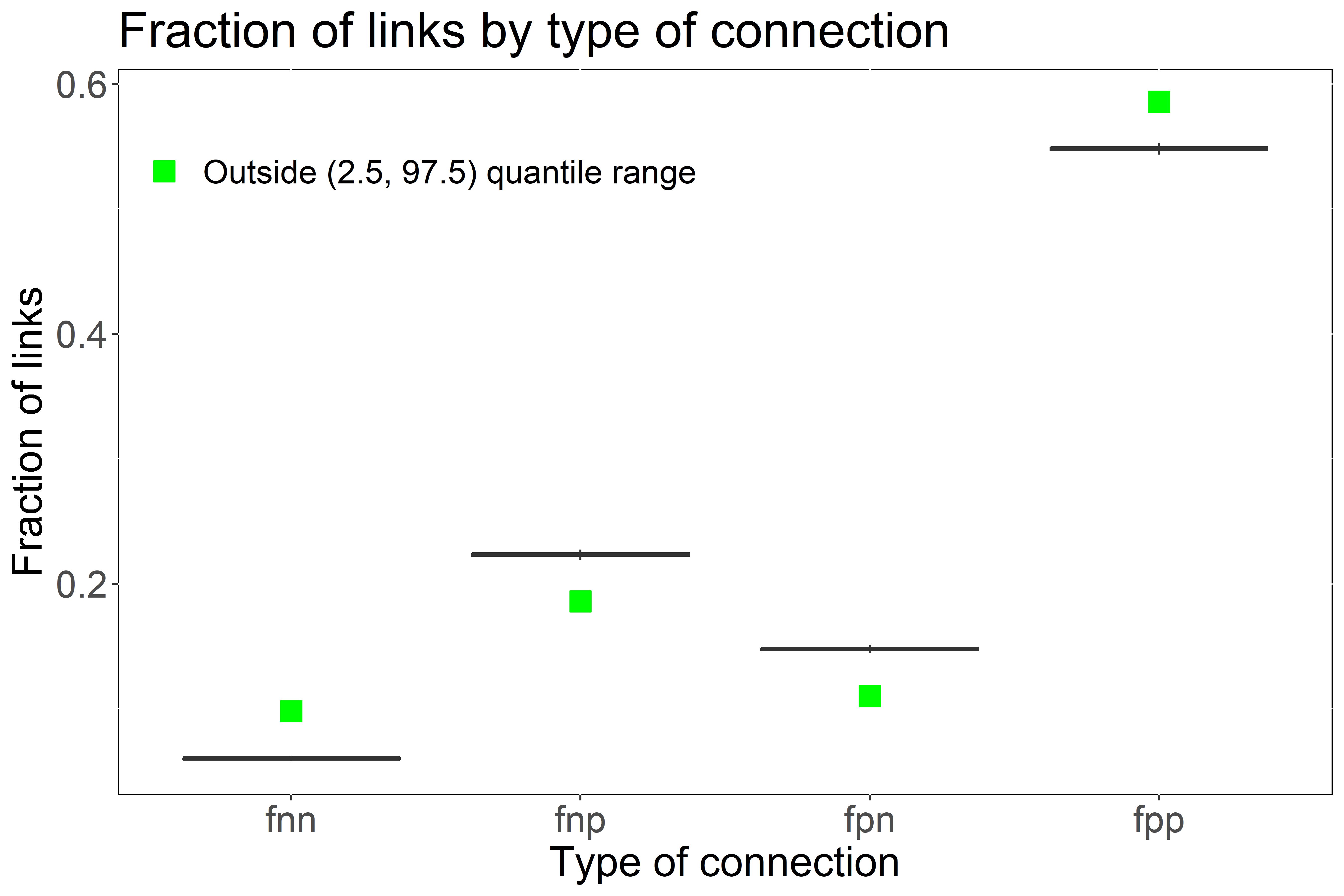}
    \caption{Results of the randomisation test in the mutual mentions network. Green squares indicate that the observed fraction of connections falls outside the lower $2.5\%$ and upper $97.5\%$ quantiles of the randomised distribution (i.e., it is unlikely to arise by chance). We denote the fraction of links between negative and negative users as \textit{fnn}, the fraction of links between negative and positive users as \textit{fnp}, between positive and negative users as \textit{fpn}, and between positive and positive users as \textit{fpp}.}
    \label{fig:boxplot_RT8}
\end{figure}

We are interested primarily in the types of connections among positive and negative nodes, which are shown in Figure~\ref{fig:boxplot_RT8}. All the observed \textit{fnn}, \textit{fnp}, \textit{fpn} and \textit{fpp} lie outside the $(2.5\%, 97.5\%)$ quantile range, meaning that connections among negative and positive nodes do not occur at random. Moreover, the same types of nodes (\textit{fnn} and \textit{fpp}) have more links between each other than what is resultant for the random distribution while opposite types of nodes (\textit{fnp} and \textit{fpn}) present smaller proportions than what would be expected at random. This is in agreement with the idea that users who show similar sentiment tend to be more connected than users who tweet opposing sentiments on the matter. Therefore, we could consider sentiment alignment as a proxy for homophily among users, i.e., ``the tendency of nodes to connect to others who are similar on some variable''~\cite{ognyanova2016network}, the variable being the aggregated sentiment score per user in our analysis.

We show that the relationships among users are not randomly distributed in the network. In fact, users who tweet similar sentiment have a tendency to be grouped together. On the network topology, this may be interpreted as a proxy for community structure in the network, where there are distinct groups of users that share the characteristics of sending (and receiving) similar sentiment to other users. Therefore, we are interested in finding community structures in the network using the sentiment scores as a proxy for how close or far apart users are. In Section~\ref{chap:communities} we analyse if users with similar aggregated sentiment scores are in the same network communities and can be clustered together.

\subsection{Finding polarised communities} \label{chap:communities}

Polarisation is commonly used to describe the division of society into groups that believe in opposite ideas. It has the potential to raise hostility and conflicts as well as occasional intergroup violence. However, it can also be a venue to challenge inequalities in societies and influence changes in policies and behaviours~\cite{smith2024polarization}. A referendum is a way to lead to policy changes; therefore, our aim in this section is to find out if there is a clear suggestion of polarisation in the online discussion during the days leading to the referendum.

The results of the randomisation process shown in Figure~\ref{fig:boxplot_RT8} suggest a polarised environment. However we can also use more standard methods to show polarised groups within the network. The tendency of people creating groups based on a shared characteristic is a common phenomenon in many social networks~\cite{newman2003social}. These groups --- the network communities --- can lead to a polarised environment in the context of a referendum, where people either vote in favour or against it~\cite{o2017integrating}. 

In this section we uncover communities based on the network structure and classify each community by observing the sentiment that users in each community send in the network. It is important to note here that our classification of users into communities is not given by the sentiment scores alone but by the network structure, where the weight of an outlink is the absolute value of the user's averaged sentiment score. Here we use the absolute value of the sentiment score for two reasons. 1) The community detection algorithms do not allow negative edge weights, and 2) the absolute value of the sentiment score is able to inform if users have a tendency of sharing more positive (or negative) language about the topic in discussion. As discussed in the previous section, users that share the same type of sentiment about the topic tend to group together, and users that share extreme language (either positive or negative) tend to be weakly connected to the opposite extreme. Therefore, ``extreme'' users (the ones sharing extreme language on average) tend to be apart from their ``extreme'' opposites on the network topology level, and the absolute average sentiment score as link weight together with the network topology constitute a strong structure to uncover polarised communities. This is a meaningful methodological improvement to polarised community detection in social networks where people (nodes) tend to group together according to their sentiment about a controversial topic. Lastly, we validate our classification with annotated data.

To check if users that share similar sentiment tend to be grouped together in the same network community and support our assumption that there exists polarisation in this social network, we apply different community detection algorithms and analyse the partition found by the algorithm that performs the best. Table~\ref{table:summ_communities} in the appendix summarises how well some commonly used community detection algorithms perform when taking the modularity as the parameter of optimisation. Modularity measures the quality of each partition in comparison to a random configuration. Consider a network with $N$ nodes and $L$ links and a partition into $n_c$ communities, each community having $N_c$ nodes connected to each other by $L_c$ links, where $c = 1,...,n_c$. The number of links within the community $c$ is defined as $L_c$, and $k_c$ is the total degree of the nodes in community $c$. To check if the local link density of the subgraphs defined by a partition differs from the expected density in a randomly wired network, we define the partition's modularity by summing the modularity over all $n_c$ communities (Eq.~\ref{eq:modularity}). The larger the value of $M$, the better the corresponding community structure~\cite{newman2003structure}. By taking the whole network as a single community we obtain $M=0$, and if we assign each node of a network to a different community, the modularity is negative.

\begin{equation}
M = \sum_{c = 1}^{n_c} \left[ \frac{L_c}{L} - \left( \frac{k_c}{2L} \right)^2 \right].
\label{eq:modularity}
\end{equation}

It should be noted that while modularity maximisation methods have been shown to have some problems (e.g., resolution limit~\cite{fortunato2007resolution}), here we are looking for two communities (the Yes and No supporters) rather than an arbitrary number, and so modularity-based methods perform well given this constraint.

By following only the modularity optimisation, the Leiden algorithm~\cite{traag2019leiden} performs best. However, the weighted Louvain~\cite{blondel2008fast} (using the absolute sentiment as the weight of a link) closely follows Leiden in modularity, and is able to find two, and only two, large communities, which are very distinct from each other and representative of the opposite sides of the discussion. Therefore, we have chosen to carry out our analysis with the two largest communities found by using the weighted Louvain method.

\begin{table}
\centering
\caption{Summary table of community clusters.}
\begin{tabular}{rrrrrrr}
\hline
  & Nodes & Links & Avg out-degree & Avg cluster coef & Density\\
\hline
C1-C1 & 2\,954 & 110\,351 & 18.373 & 0.376 & 0.006\\
C2-C2 & 463 & 5\,165 & 3.752 & 0.378 & 0.008\\
C1-C2 & 861 & 2\,401 & 1.777 & 0 & 0.002 \\
C2-C1 & 897 & 4\,955 & 2.881 & 0 & 0.003\\
\end{tabular}
\label{table:summ_CC}
\end{table}

Table~\ref{table:summ_CC} is a summary of the properties of each community. We denote C1-C1 as the subgraph built with users in Community 1 only, i.e., effectively C1 only. Similarly, C2-C2 is effectively the C2. C1-C2 is the subgraph containing the links from C1 to C2, and C2-C1 is the subgraph containing the links originating in C2 and ending in C1. Community 1 (C1) is approximately six times the size of Community 2 (C2) on number of nodes (that is comparable to the number of nodes in the Yes and No communities observed for the Irish Marriage Referendum, seen in Appendix~\ref{chap:IMR}). Although C1 has a higher average out-degree than C2, C2 is more tightly connected as it presents a slightly higher average cluster coefficient and higher density. In contrast, both C1-C2 and C2-C1 present lower average out-degree and are less tightly connected than their counterparts since both average cluster coefficient and density are smaller. We also observe that users in C2 mention C1 more frequently than the opposite. This may be due to the fact that C2 is much smaller in number of nodes than C1.

\begin{figure}[h] 
    \centering
    \includegraphics[width=\textwidth]{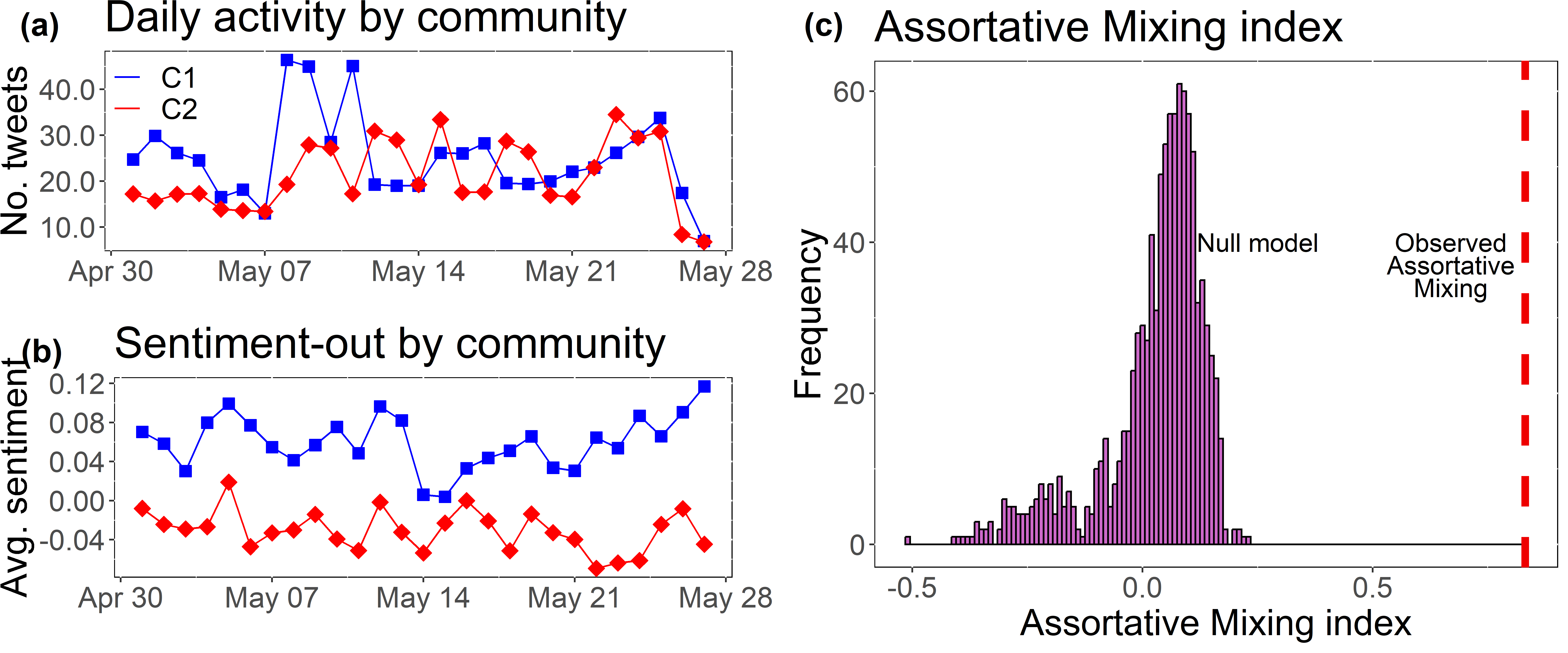}
     \caption{(a) Daily average number of tweets by community over time. C1 (Yes community) in blue and C2 (No community) in red. (b) Average sentiment-out of communities over time analysed. C1 presents a higher average sentiment-out over time than C2. (c) Monte Carlo simulation (histogram) of the assortative mixing index together with the observed assortative mixing index. The purple bars are the resultant correlation distribution obtained after $1\,000$ simulations, and the red dashed line represents the observed index ($0.83$ --- high assortative mixing).}
    \label{fig:summary_comm}
\end{figure}

To classify the communities into the Yes community and the No community, we check the average sentiment-out of each of the communities over time (Figure~\ref{fig:summary_comm}~(b)). We classify C1 as the Yes community --- it presents a higher average sentiment-out over time --- and C2 as the No community, since it presents a lower average sentiment-out over time. Figure~\ref{fig:summary_comm}~(a) shows that there is no significant difference between each community activities over time. Figure~\ref{fig:summary_comm}~(c) reassures that the assortative mixing in our network is significant. The simulation process is as follows:

\begin{enumerate}
\item Randomly assign a community membership (1 or 2) to each user in the mutual mentions network with replacement. Here, we use sampling with replacement to make sure the probability of selecting either community membership remains the same in future draws. This procedure also keeps the network topology intact.
\item Calculate the assortative mixing index $r$ for the network with new community assignments.

\begin{equation}
    r = \frac{\sum\limits_{i} e_{ii}- \sum\limits_{i} a_ib_i}{1 - \sum\limits_{i} a_ib_i},
\end{equation}

where $e_{ii}$ is the observed fraction of links that connect two nodes which both have value $i$, $a_i$ is the probability that a link has origin in a node with value $i$, and $b_i$ is the probability that a link has as destination a node with value $i$~\cite{coscia2021atlas}.

\item Repeat steps 1-2 $1\,000$ times to create the null model for where the assortative mixing index could lie if there was no connection between the network structure and the community assignment.
\end{enumerate}

The observed assortative mixing ($0.83$) is high on a scale that ranges from $-1$ (disassortative) to $1$ (complete assortativity)~\cite{coscia2021atlas}, while the simulations lie between $(-0.3, 0.2)$. 

\begin{figure}[h] 
    \centering
    \includegraphics[width=13cm]{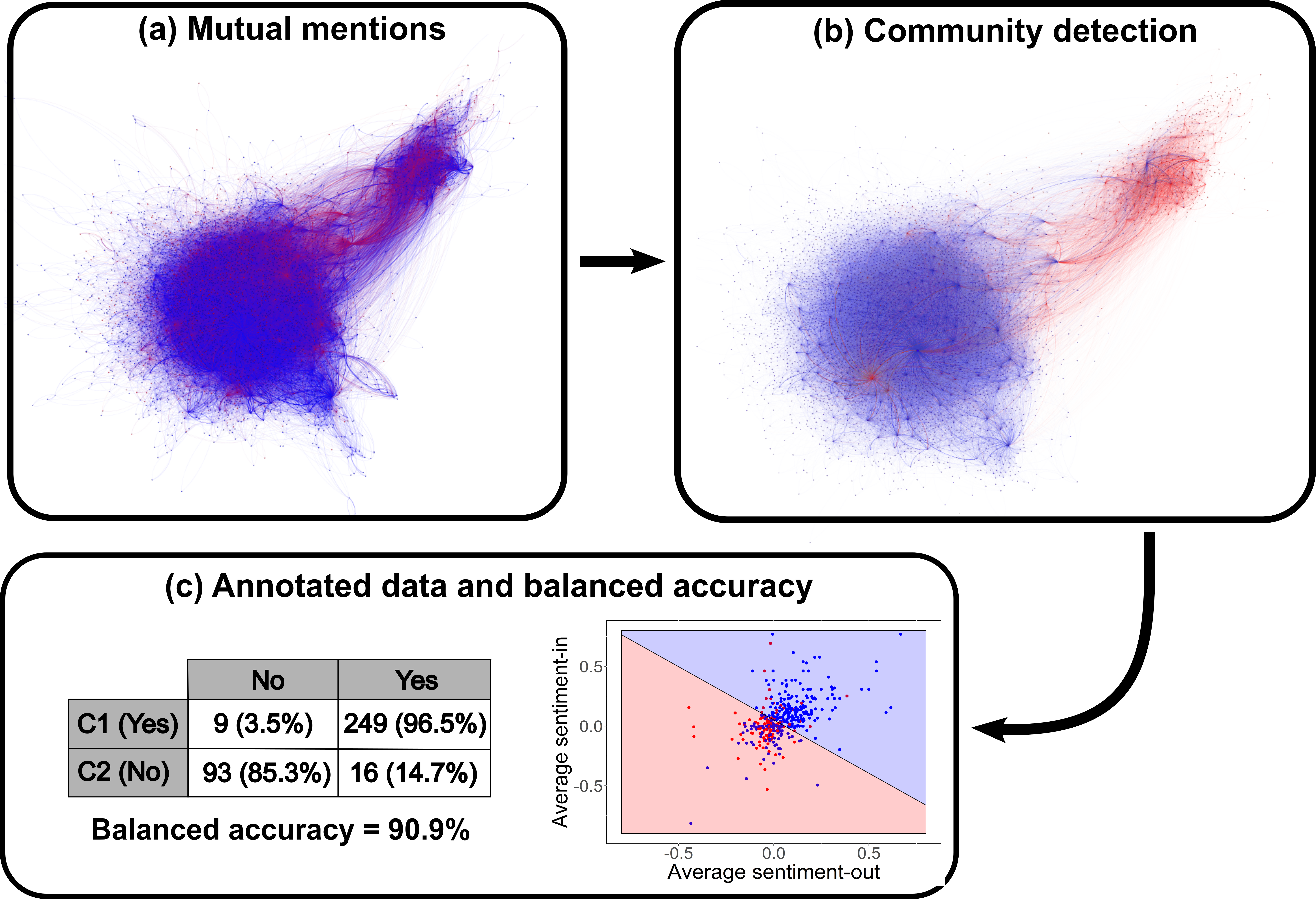}
    \caption{(a) Mutual mentions network representation with nodes and links coloured by the sentiment (positive or negative) of each individual; (b) Mutual mentions network representation with nodes and links coloured by the community of each individual, after applying the Louvain community detection algorithm; (c) To check accuracy, we hand-classified over $10\%$ of the users in the mutual mentions network. The majority of the Yes supporters lie on top of the Yes community, and the opposite is also true. The balanced accuracy\protect\footnotemark of our method of community identification is $90.9\%$.}
    \label{fig:comm_detection_RT8}
\end{figure}

\footnotetext{Balanced accuracy is a metric used to assess the performance of a classification method when the classes are imbalanced. Balanced accuracy = $\frac{\textrm{(True Yes users rate + True No users rate)}}{2}$.}

Figure~\ref{fig:comm_detection_RT8} shows how the community detection method accounts not only for the user sentiment scores (Figure~\ref{fig:comm_detection_RT8}~(a)) but also for the social structure. In Figure~\ref{fig:comm_detection_RT8}~(b) the distinction between the two communities is clear and there is a suggestion of polarisation, where each ideological community tends to be apart from each other, with little conversation in between. To determine the performance of our method of classification, we manually classify a stratified random sample of $10\%$ of users in Community 1 and $10\%$ of users in Community 2 as either Yes or No supporters (please refer to Appendix~\ref{chap:appendix} for more information). We check this distribution against the community assignment, which can be seen in  Figure~\ref{fig:comm_detection_RT8}~(c). This shows that the majority of manually-classified No supporters are in the No Community (C2) and the majority of the Yes supporters lie in the Yes Community (C1). From the confusion matrix we see that we have good agreement between our method of classification and the reality (based on data annotation). The overall balanced accuracy of the method is $90.9\%$, which is considered very good~\cite{brodersen2010balanced, wei2013role}. This result is more accurate than the one obtained by using a combination of the mentions and the followership networks (Appendix~\ref{chap:follower}), and more accurate than the method using hashtags for user classification (Appendix~\ref{chap:hashtags}).

To summarise our classification method, the following are the steps needed:

\begin{enumerate}
\item Build a conversation network among highly active users.
\item Add weights for links between users where Weight = $|$Averaged out-sentiment of the node where the link starts$|$.
\item Apply well-understood community classification methods on the network, using weights when allowed.
\item Check which classification method returns the best results.
\item (If needed) Apply clustering method (e.g., k-means), where the parameters for clustering are the average in-sentiment and average out-sentiment of users, to cluster communities together. Appendix~\ref{chap:IMR} shows the classification of the Irish Marriage Referendum network, which needs this step.
\item Classify each community as prone to the yes-side or the no-side of the debate given the sentiment their users send to others.
\item Validate the classification method with annotated data.
\end{enumerate}

\begin{figure}[h]
    \centering 
    \includegraphics[width=0.9\textwidth]{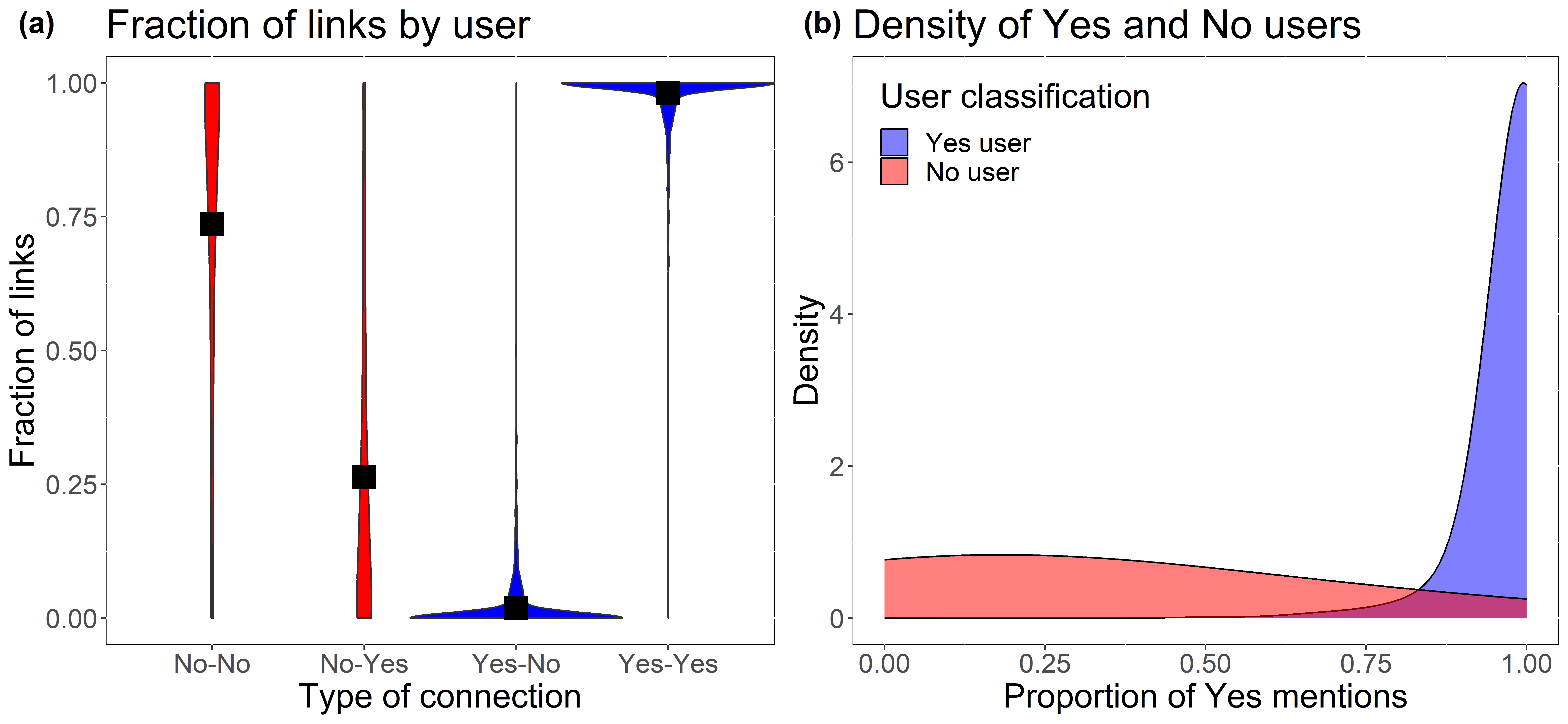}
    \caption{Fraction of connections between users in the two community clusters. (a) Violin plot for the distribution of types of links by user, aggregated by community-type. The black squares represent the mean for each distribution. (b) Density plot for the distribution of Yes links by user, aggregated by community type.}
    \label{fig:fraction_links_comm}
\end{figure}

To reinforce our point that communities in the Repeal the $8^\textrm{th}$ mentions mutual network are polarised, Figure~\ref{fig:fraction_links_comm}~(a) shows the distribution of fractions of types of links by each user in the network. We observe that users tend to communicate with others in the same community and interact less with users in the opposite community. Figure~\ref{fig:fraction_links_comm}~(b) shows that it is unlikely that Yes users will mention a No user (as very few of them have links to No users). The opposite is still true, but to a lesser extent as there is a small number of No users who have links to the Yes side of the referendum. Given that we know, with a high degree of accuracy, the community structure and also how communities share information, we can analyse how information spreads inside and between communities. In the following section we do this by reconstructing the cascades of information diffusion.

\subsection{Information diffusion within and cross-community} \label{chap:cascades}

Information diffusion can be represented as a cascade of events. A cascade allows us to trace the pattern of diffusion from the user that initiated the process (the seed) through a population~\cite{goel2016structural}. It has been hypothesised by some that cascade behaviours are hard to predict~\cite{martin2016exploring}. However, insights as to how the network's structure impacts the spreading of content and the mechanism by which it spreads can still be obtained~\cite{keating2023generating, dave2011modelling, peters2008modelling}. We are particularly interested in how information spreads within and between communities~\cite{weng2013virality}.

To this end we employ a popular cascade reconstruction method proposed by Goel et al.~\cite{goel2016structural}, in which they use the temporal sequence of retweets and the network structure to reconstruct the likely cascade structure. Even though this is a mechanistic rule-based method, it attained very reasonable performance in the original paper and in a previous application~\cite{gleeson2020branching}. The process can be broken down into four steps: 1) bucketing Retweet-IDs, 2) ordering chronologically, 3) parent attribution, and 4) cascade-ID assignment. 

Following this process, we first compare every tweet in our collected dataset (but excluding tweets without mentions) to create Retweet-IDs for each unique cleaned text (lower case, without mentions and hashtags), then order the tweets in the Retweet-IDs according to the time they were posted. There are $114\,739$ original tweets that were never retweeted. Of those, $29\,108$ ($25.37\%$) were written by Yes users, $6\,305$ ($5.49\%$) were tweeted by No supporters, and $79\,326$ ($69.14\%$) by users that were not assigned an ideological community. 

\begin{figure}[h] 
    \centering
    \includegraphics[width=13cm]{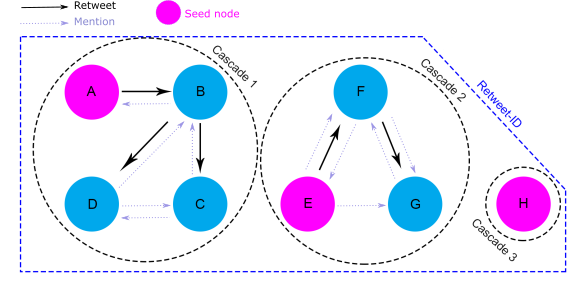}
    \caption{Method of finding parents and children using mentions in the network. Nodes are named according to the time where they (re)tweeted the content, i.e., A tweeted the content first in our timeline, while H was the last one to tweet that text. Pink nodes are the seeds of their cascades. Dashed grey arrows specify who mentions whom in the network, and full black arrows show the parenting assignment.}
    \label{fig:finding_parents}
\end{figure}

\begin{figure}[h]
     \centering
     \includegraphics[width=0.8\textwidth]{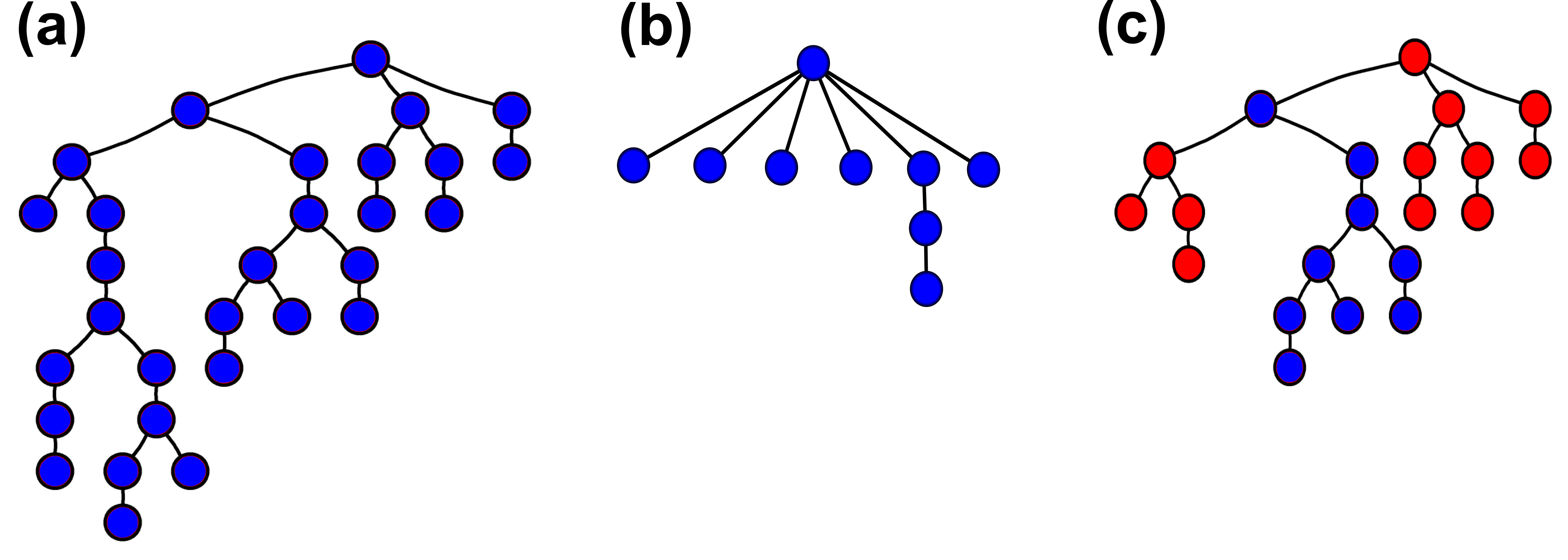}
    \caption{Examples of cascade-trees built using the mentions network. Users in the Yes community are represented as blue nodes, while those in the No community are red. (a) A Yes-sided cascade that presents a viral behaviour, in which many users independently pass the information ahead. (b) A Yes-sided cascade that presents a broadcast behaviour, in which many users copy the content from one big source. (c) A cascade presenting a mix of broadcast and viral behaviours composed by both types of users (Yes and No supporters).}
    \label{fig:trees}
\end{figure}

Next, we find who, among the user's mentions (here, we again use the whole original dataset so that we avoid biases), is the most likely to have been retweeted from, for each retweet in each Retweet-ID (Figure~\ref{fig:finding_parents}). We assign a parent (A) to a mentioned user's (B) retweet if they are the user who most recently could have introduced the content into B's timeline (i.e., A tweeted the same content before B and B mentions A at some point in our network). If no such mention exists, we treat B as a root of an independent cascade, i.e., they posted that content independently, starting a new tree (e.g., users E and H in Figure~\ref{fig:finding_parents}). From the original $15\,340$ Retweet-IDs containing at least two tweets each (the original tweet and at least one retweet), we retrieved $24\,983$ cascades when applying the method of parent-children attribution. Examples of trees retrieved from this process are shown in Figure~\ref{fig:trees}, where blue nodes are users in the Yes community and red nodes are users in the No community. 

\begin{figure}[h]
     \centering
    \includegraphics[width=0.9\textwidth]{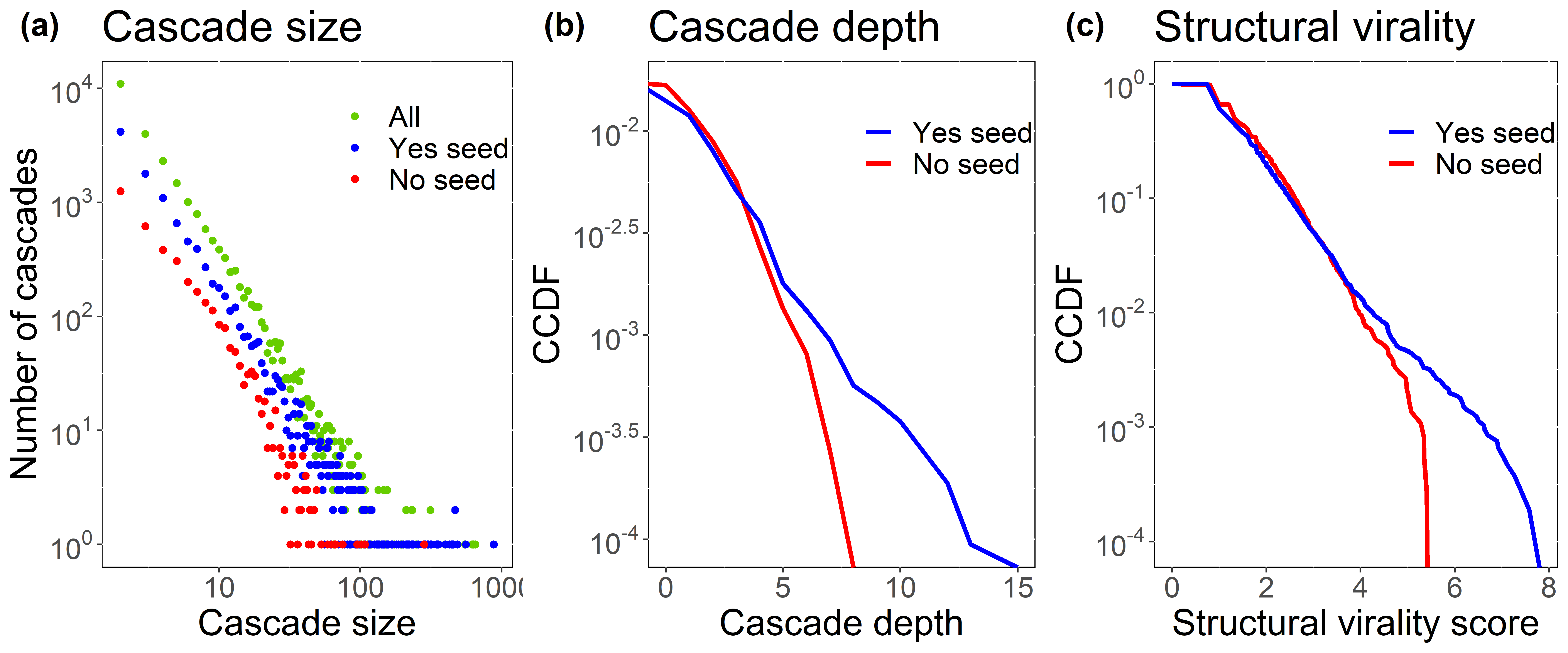}
    \caption{(a) Distribution of cascades sizes by seed community. (b) CCDF of maximum cascade depth by seed community. (c) CCDF of structural virality scores by seed community.}
        \label{fig:seed_sizes}
\end{figure}

To analyse how the information spreads across the same ideological community and across different communities, we classify the cascade trees as Yes seeded or No seeded according to which community the first user who tweeted the content (the seed) belongs to. Figure~\ref{fig:seed_sizes} shows that the distribution of cascade sizes is heavy-tailed, that is, a large number of cascades are small (from two users), while only a small number of cascades are large --- which complies with previous works~\cite{goel2016structural, lerman_information_2016, vosoughi_spread_2018, juul_comparing_2021}. Of the $24\,983$ cascades retrieved (Figure~\ref{fig:seed_sizes}~(a) green curve), $10\,631$ ($42.5\%$) are Yes-seeded (Figure~\ref{fig:seed_sizes} blue curves) and $3\,774$ ($15.1\%$) are No-seeded (Figure~\ref{fig:seed_sizes} red curves). The remaining ($42.3\%$) have seeds unclassified into communities --- these could be unknown or could belong to one of the ideological communities but were filtered out when we chose to analyse only the strongest connected component of the mutual network and the two largest communities uncovered by the weighted Louvain method. The largest Yes-seeded cascade contains 884 users, and the largest No-seeded cascade contains 283 users. The disparity between the sizes of cascades may be due to the fact that there are fewer No supporters than Yes supporters in our network and also because the Yes-seeded cascades are more abundant than the No-seeded ones, therefore we have a smaller probability of finding large cascades amidst the No-seeded cascades. Depths and virality reach greater values on the Yes-seeded cascades (Figures~\ref{fig:seed_sizes} (b) and (c)). The Yes-seeded cascades reach depths as high as $15$, while the No-seeded cascades do not surpass eight steps --- again, this might be an effect of the disparity between the number of each type of cascade. The Yes-seeded cascades present slightly more viral behaviour (maximum virality of $7.8$) in comparison with the No-seeded cascades (which reach virality of $5.4$ the highest). The distributions observed for cascade size and cascade depth are comparable to the ones found in previous works on social media networks, however the structural virality is slightly lower when compared to other online discussions~\cite{goel2016structural, vosoughi_spread_2018, juul_comparing_2021}, suggesting a more broadcast behaviour --- many users sharing information from a small amount of key users in the discussion. See Appendix~\ref{chap:appendix_cascades} for more details on cascade scores.

\begin{figure}[h] 
    \centering
    \includegraphics[width=0.85\textwidth]{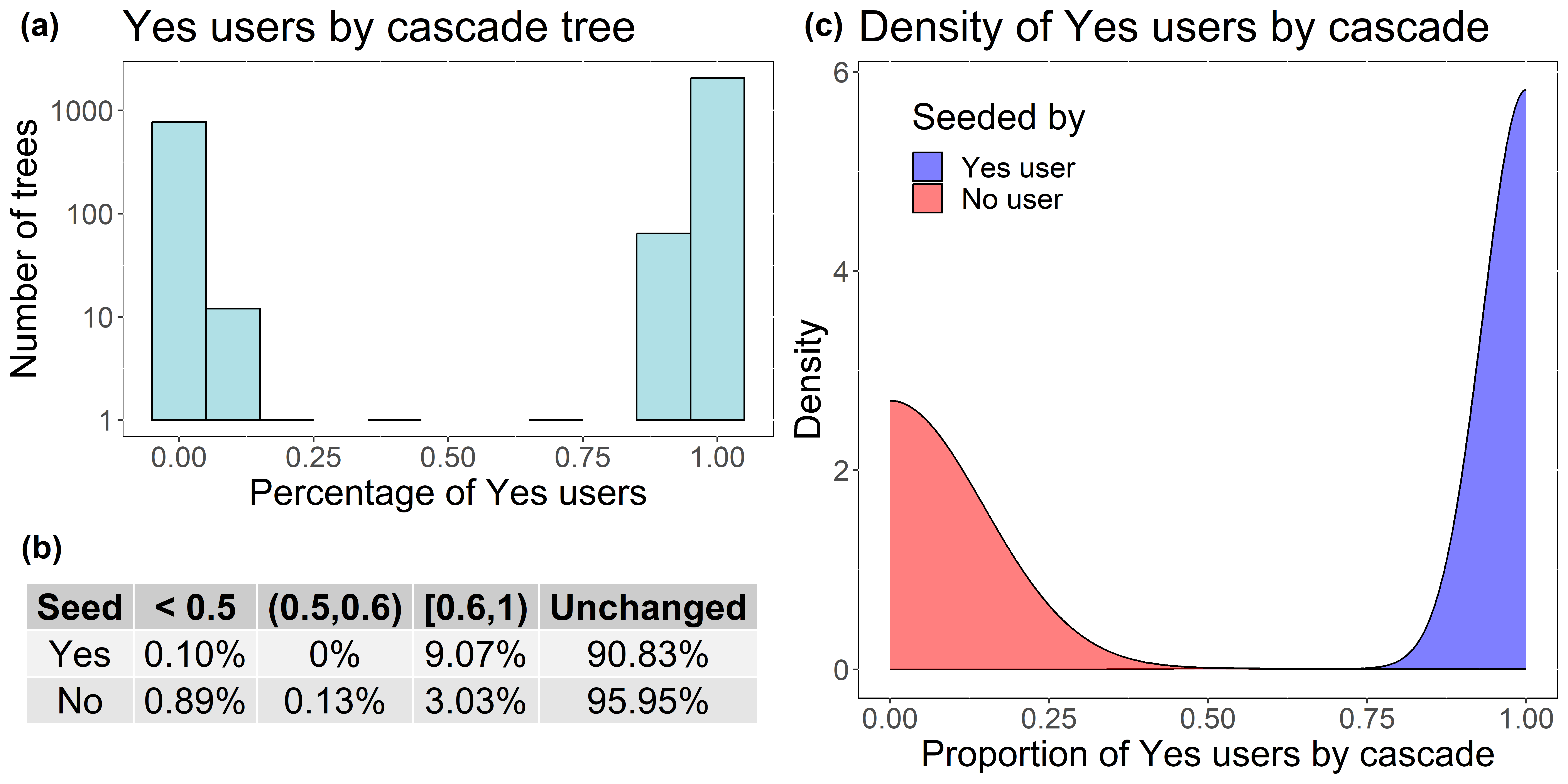}
    \caption{(a) Proportion of Yes users by cascade tree (of the cascades that have 10 or more classified users). Note that the y-axis is on a log-scale. (b) Behaviour of information diffusion by seed community (of the cascades that have 10 or more classified users). (c) Density of Yes users by cascade tree, coloured by cascade seed type (of all cascades).}
    \label{fig:prop_yes_cascade}
\end{figure}

To determine how information spreads between and inside communities, of the $2\,918$ cascades that have 10 or more classified users --- the ones classified into one community or the other --- only two ($0.07\%$) have a percentage of Yes users between $(0.25,0.75)$, which means that the majority of the cascades tend to present $75\%$ or more users of the same ideological community (Figure~\ref{fig:prop_yes_cascade}~(a)). Or, in other words, information tends to diffuse inside communities, and less frequently between communities. We also observe (Figure~\ref{fig:prop_yes_cascade}~(b)) that $90.83\%$ of the information that starts with a Yes seed remains in the Yes community, and over $95\%$ of the information that starts with a No seed remains in the No community. The table shows the proportion of cascades that have less than $50\%$ of the users in the same community as the seed (when we consider that the community where the information spread changed) between $50\%$ and $60\%$, between $60\%$ and $100\%$,  and the proportion of cascades where all users are in the same community as the seed (Unchanged). It shows that only $0.10\%$ of the cascades starting with a Yes user entered the No community, and only $0.89\%$ of the cascades starting on the No side spread to the Yes side of the discussion. Figure~\ref{fig:prop_yes_cascade}~(c) shows the proportion of Yes users by cascade grouped by the type of seed of each cascade, now considering every cascade retrieved. We observe that the majority of No-seeded cascades present a low density of Yes users, while Yes-seeded cascades tend to have a high density of Yes users. 

Therefore, we show that information tends to spread mainly inside the same ideological community, and less frequently we observe information diffusion between the opposite ideological communities. Those results are comparable to the ones obtained by retrieving cascades using the followership network (Appendix~\ref{chap:follower}) --- which is known to be successful~\cite{goel2016structural, vosoughi_spread_2018} --- if not more interesting as it avoided splitting up cascades into smaller ones when a parent was not found. We observe that users do not always follow people they retweet from, but they likely interact with them at some point, be it a reply or a mention to the content they post. Also, the mentions network tends to have more complete information in the sense that the followership is recovered only for the exact time when the data is being mined, which might take place years after the phenomenon we are studying occurred --- as in our case, a four-year gap --- and, so, users might have stopped following others or might have deleted their accounts (in which case they would still be present in our mentions network, as mentions are not deleted by deletion of accounts).

We therefore show that the vast amount of information shared between two communities never leaves the host community, in effect there is an echo chamber of information. This also stands in contrast to Figure~\ref{fig:fraction_links_comm}~(a), where we noted that there was indeed a group of users who mentioned users in the other community (No did mention Yes), but these interactions did not result in retweets. 


\section{Discussion and conclusion} \label{chap:conclusion}

The understanding of social interactions, polarisation of ideas, and the spread of information plays an important role in society. We used Twitter data to identify opposing sides of the Irish Abortion Referendum of 2018 debate and to observe how information spreads between these groups in our current polarised climate. We built a sentiment-based mentions network from the tweets concerning the debate to detect polarised communities --- communities dominated by one type of user (Yes or No supporter) --- with high balanced accuracy ($90.9\%$). We showed that we were able to do it without the need for the followership network, which requires a time-consuming gathering process, and by using this simpler approach we achieved higher accuracy than previous work~\cite{o2017integrating}.

These communities and over $31\,000$ retweets formed the basis of our analysis of information diffusion, which showed that information tends to spread heavily inside the same ideological community and less frequently between communities. This means that users tend to communicate mainly with those with whom they share an ideology and less with people that have opposite opinions regarding the abortion debate in Ireland. This provides a valuable methodology for extracting and studying information diffusion on large networks by isolating ideologically polarised groups and exploring the propagation of information within and between these groups. Once again, we did not make use of the followership network as seen in previous studies~\cite{goel2016structural, vosoughi_spread_2018, juul_comparing_2021}. Our results show that by using the conversation network instead of the followership network for parenting attribution, we avoid splitting the Retweet-ID cascades into many smaller cascades by not finding followership links between users.

Some of the limitations of our analysis are due to the data gathering process as we may not be able to retrieve all the original data regarding the referendum. However, the network that we created is large, well-connected and robust for our purposes. Another concern with the data gathering method is the recent changes to the Twitter API, as Twitter no longer gives researchers access to data via their academic API. Therefore, trying to collect Twitter data might incur unwanted extra expense. Given that our method used significantly less data than previously published methods to arrive at nodal stances, with no loss in accuracy, this would be a preferred method to be used for node classification and information cascade building. Also, as we have seen, a sentiment lexicon might struggle to catch all the nuances of double negatives and irony~\cite{silgeTextMiningTidy2017}. However, we mitigated this weakness by (a) aggregating on a user-level so if a user has a few sarcastic tweets, the majority of their tweets still have straightforward language, which prevails when aggregated, and (b) by performing multiple randomisation tests to show that sentiment does indeed provide a signal that helps explain the network topology, where the sentiment that a user sends and receives is highly correlated, along with the connectivity patterns between Yes and No supporters. Finally, the use of a community detection method that may have its own limitations~\cite{fortunato2007resolution} proved to provide a valuable and accurate clustering of the users into two polarised groups. Another limitation that is worth discussing is the possibility of off-channel diffusion of information, which is not captured by our cascade inference method. The same piece of content might spread via other channels such as Facebook and Youtube, and two users that are connected on other platforms might independently post the same content on Twitter after seeing it on other platforms, and in this case our method mistakenly treats a single diffusion tree as two disjoint events~\cite{goel2016structural}.

In future work we plan to show how the empirical cascades compare to models of information diffusion. Given that we have found that information rarely spreads between communities on this topic, we are interested in finding 1) which of the popular models of information diffusion best capture this spread and 2) whether the network structure sufficiently mediate the observed spread of information between the communities or whether we require information on community specific content-spreading ``attractiveness''. We will also use centrality measures to find the most influential users in our polarised environment and assess their performance when applied to a polarised social network. Furthermore, we would like to explore other methods for node classification in future work using machine learning algorithms, such as knowledge graphs and text embedding, and potentially extend our initial classification of very active nodes to more peripheral nodes in the discussion using label propagation. We would also like to explore other methods of cascade inference in the future, such as NetInf~\cite{gomez2012inferring} and other statistically based methods, and assess how well these methods match each other on empirical datasets and how well they recreate known cascade structures from simulated data.


\section*{Acknowledgements} \label{chap:acknowledge}

The authors are very thankful to Laura O'Mahony for the help with data gathering, Emily O'Sullivan and Rory McCarty for the great work with data annotation, and Samuel Unicomb for the valuable and constructive feedback.

This publication has emanated from research conducted with the financial support of Science Foundation Ireland under Grant number 18/CRT/6049. For the purpose of Open Access, the authors have applied a CC BY public copyright licence to any Author Accepted Manuscript version arising from this submission.

\newpage

\appendix

\section{Extra details} \label{chap:appendix}

\subsection{Mentions network creation --- an illustrative example}

In this appendix we provide an illustrative example of how the mentions network was created. In our example, we consider two users: User 1 and User 2. Both User 1 and User 2 use tracked hashtags in their tweets, they are in the strongest connected component of the network, and they mention each other at least once. Therefore, their tweets concerning the Repeal the $8^\textrm{th}$ Referendum are in the mutual mentions network to be analysed. Each tweet is assigned a sentiment score by summing the re-scaled positive with the re-scaled negative scores. The re-scaling depends on the maximum positive score and the maximum absolute negative score observed in a single tweet in the mutual mentions network. In our example, the maximum positive score is 50 and the maximum absolute negative score is 40 (which, divided by 5, the maximum allowed score, are 10 and 8, respectively). The resulting sentiment by user is the average score of all their tweets in the network. A positive sentiment score means that, on average, the user tweets using more positive words than negative ones.
 
\begin{figure}[H] 
    \centering
    \includegraphics[width= 14cm]{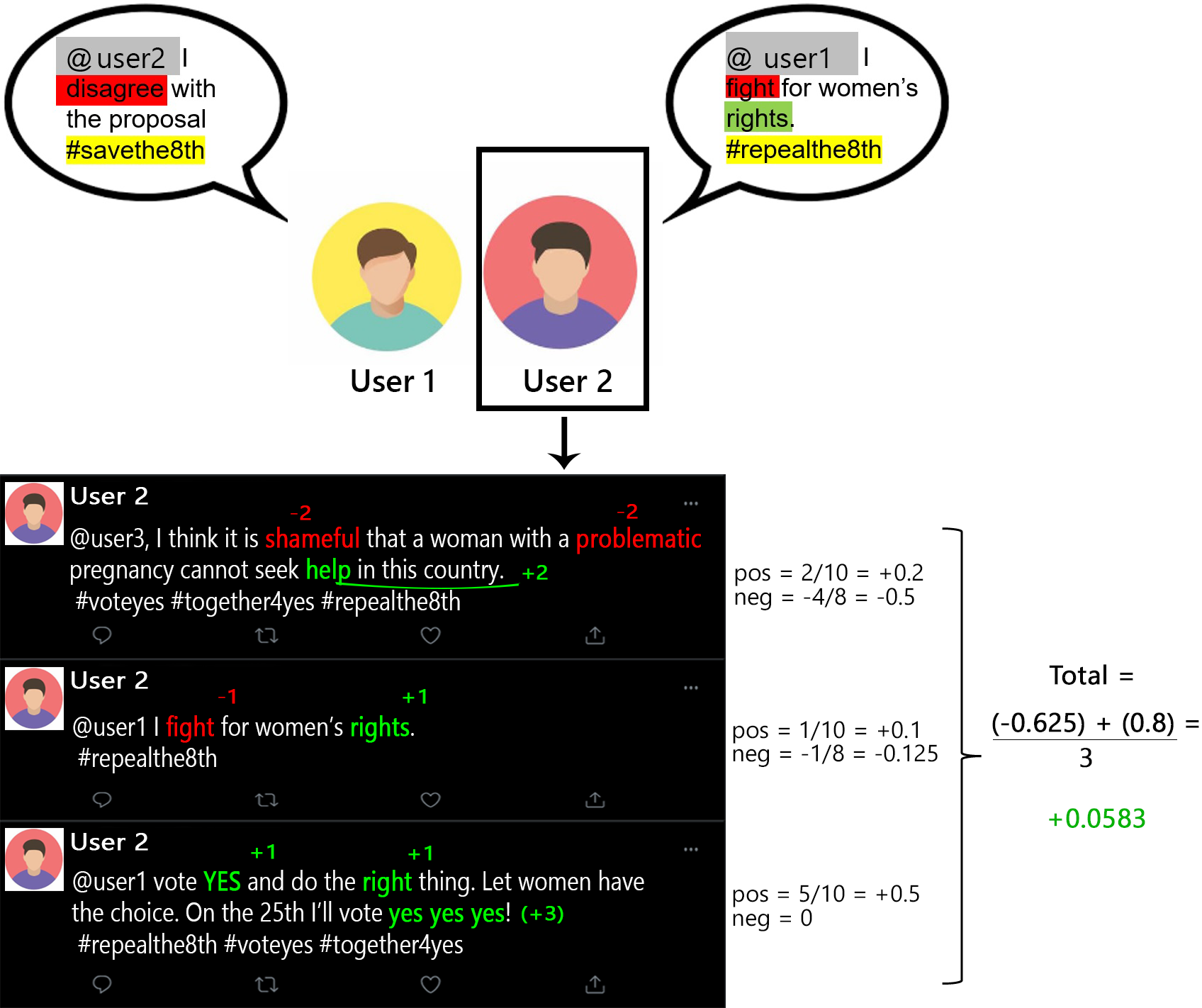}
    \caption[Example of how the mentions network was built]{Example of how the mentions network was built and how  we assign sentiment scores. Words are detected as positive (green) or negative (red), and given a score. The zoom in User 2's tweets exemplifies how the re-scaling and averaging were done. In this example, we consider that the highest positive score in a single tweet in the network is +50, and the lowest negative score in a single tweet in the network is -40. Note: the texts and usernames in this example are merely illustrative.}
    \label{fig:sentiment}
\end{figure}

\subsection{Community detection algorithms}

To find the best community detection algorithm for our case (Section~\ref{chap:communities}), we compare some of the most used detection algorithms based on modularity. The results are shown in Table~\ref{table:summ_communities}. We call ``significant'' the communities which had 20 nodes or more, selecting only those to compose our final mutual mentions network. The table shows that although the Leiden~\cite{traag2019leiden} detection method achieved the highest modularity score, Louvain~\cite{blondel2008fast} follows it closely and retrieves only two large communities, which show to be very representative of the polarised communities we are looking for.

\begin{table}
\centering
\caption{Summary table of community detection algorithms. ``Significant'' communities are the ones containing 20 or more nodes.}
\begin{tabular}{lrl}
\hline
Algorithm & Modularity & Number of communities\\
\hline
Leiden & 0.298 & 13 communities, 11 significant\\
Louvain & 0.280 & 217 communities, 2 significant\\
Spinglass & 0.273 & 18 communities, 8 significant\\
Fast greedy & 0.246 & 523 communities, 7 significant\\
Walktrap & 0.195 & 49 communities, 7 significant\\
Leading eigen & 0.191 & 5 communities, all significant\\
Infomap & 0.071 & 552 communities, 31 significant\\
Label propag & 0.003 & 39 communities, one containing almost all nodes\\
\end{tabular}
\label{table:summ_communities}
\end{table}

\subsection{The manual classification method}

In Section~\ref{chap:communities} we assess the accuracy of our method of users classification by manually classifying $20\%$ of the users in the mutual network after community detection. We manually classify a stratified random sample of $10\%$ of users in Community 1 and $10\%$ of users in Community 2 as either Yes or No supporters. Three labelers were involved in this process, one of the authors and two research interns (see Acknowledgements). The labelers were not provided with the tweet scores to avoid biases, and their classification was based solely on the texts and hashtags the users sent to others. The manual classification was based on all tweets that the user being analysed sent to others in our network. If the language they used is indicative that they supported the Yes side of the debate, they were classified as a Yes voter, and likewise a user that sent tweets suggesting support to the No side was classified as a No voter. As an example, User 2 in our illustration in Figure~\ref{fig:sentiment} clearly supports the Yes vote, and would be manually classified as a Yes voter. Thirty-eight out of the 780 users analysed ($5\%$) did not present enough information for classification as Yes or No voters, and were excluded from the accuracy assessment.

\section{Hashtags: data collection and community classification method} \label{chap:hashtags}

This section is a further look on the Repeal the 8$^\textrm{th}$'s hashtags scraping and dynamics. 

\subsection{Data collection}

For data collection, we first gathered tweets containing either \#repealthe8th or \#retainthe8th, allegedly used by opposite sides of the discussion. From the initial dataset, we observed that the users in our scrapped dataset would also often use the hashtags \#loveboth, \#together4yes, and \#savethe8th, therefore the second step in the data scrapping also included these hashtags. This process added $441\,399$ ($64\%$ of the total) tweets and $81\,024$ ($42.9\%$ of the total) users to the dataset. 

Table~\ref{tab:hashtags} shows the 15 most used hashtags in the $688\,945$ tweets gathered and their representativeness when considering every unique hashtag in the dataset. Those 15 hashtags together account for $75\%$ of all hashtags, and \#repealthe8th, \#together4yes, and \#savethe8th account for $51.04\%$ of all the hashtags in our dataset. Some hashtags in the list are not specific to the Repeal the 8$^\textrm{th}$ Referendum, such as \#hometovote, \#voteyes, \#repeal, \#ireland, \#yes, and \#voteno, therefore could not have been used to scrape more tweets. Those hashtags together represent $15.45\%$ of all hashtags in the dataset. We therefore show that the chosen hashtags used to gather data on the Twitter API are representative of the referendum's online discussion.

\begin{table}[H]
    \centering
    \caption{The 15 most common hashtags in the Repeal the 8$^\textrm{th}$ dataset and their representativeness when considering every unique hashtag in the scrapped tweets. Rows in green show the hashtags we have used to gather data. Here, the percentage is normalised on the 15 hashtags only.}
    \begin{tabular}{lrr}
    \hline
        Hashtag & Number & Percentage \\
        \hline
        \rowcolor{green}
         repealthe8th & 248\,107 & 41.95\% \\
         \rowcolor{green}
         together4yes & 112\,134 & 18.96\% \\
         hometovote & 66\,168 & 11.19\% \\
         \rowcolor{green}
         savethe8th & 42\,204 & 7.14\% \\
         8thref & 24\,102 & 4.07\% \\
         voteyes & 22\,998 & 3.89\% \\
         togetherforyes & 18\,326 & 3.10\% \\
         repeal & 9\,579 & 1.62\% \\
         ireland & 9\,106 & 1.54\% \\
         yes & 7\,254 & 1.23\% \\
         voteno & 6\,781 & 1.15\% \\
         referendum2018 & 6\,555 & 1.11\% \\
         \rowcolor{green}
         loveboth & 6\,292 & 1.06\% \\
         repealedthe8th & 6\,104 & 1.03\% \\
         lovebothvoteno & 5\,694 & 0.96\% \\
    \end{tabular}
    \label{tab:hashtags}
\end{table}

\subsection{Using hashtags to classify users into polarised communities}

Bruns et al., Conover et al., and Chen et al.~\cite{bruns2017australian, conover2011political, chen2021polarization} showed that hashtags can be used to classify Twitter users as taking part on specific discussions. Darwish~\cite{darwish2019quantifying} showed that those who supported and opposed (on Twitter) the confirmation of Kavanaugh to the US Supreme Court were generally using divergent hashtags. Recuero et al.~\cite{recuero2015hashtags} and Chagas et al.~\cite{chagas2022far} used hashtags to classify users into one side of the debate being analysed. In our work, we showed that we can use a combination of the network structure and sentiment analysis to do that classification with a high balanced accuracy of $90.9\%$. The aim of this section is to analyse how the hashtag classification performs compared to the ground truth --- manually classified users --- and to our own method. 

To classify a user into either a Yes voter or a No voter by analysing hashtags, we firstly classify each tweet in our dataset into Yes or No depending on the hashtags contained in the text (here, we consider \#repealthe8th and \#together4yes as indicative of the Yes side, and \#savethe8th, \#loveboth and \#retainthe8th as proxy for the No side). We then classify each user by the most frequent side of their tweets. However, it is possible (and even common) that a person uses different --- sometimes opposite --- hashtags in the same tweet and across their tweets. The hashtag \#repealthe8th is often used as a neutral hashtag to the discussion, as well as by the yes side. We also observe people using the hashtag \#repealthe8th together with \#savethe8th, \#loveboth, and \#retainthe8th, allegedly no-side hashtags. Furthermore, it is possible that people use opposite hashtags to call attention to the opposite side of the debate.

\begin{table}[H]
\centering
\caption{Confusion matrix hashtag-based user classification and manual classification. The balanced accuracy is $88.88\%$.}
\begin{tabular}{c|ccc}
\hline
 Hashtag/Manual & No & Yes\\
\hline
No & 84 (91.3\%) & 8 (8.7\%)\\
Yes & 21 (13.5\%) & 134 (86.5\%)\\
\end{tabular}
\label{tab:cross_hash_support}
\end{table}

Table~\ref{tab:cross_hash_support} shows the confusion matrix between the hashtag-based classification and the manual classification as described in Section~\ref{chap:communities}. The balanced accuracy by using this method of classification is $88.88\%$, a bit smaller than but comparable to the balanced accuracy of our method of classification. Table~\ref{tab:cross_hash_memb} shows the confusion matrix between our method of classification and the hashtag-based user classification. The balanced accuracy between the two methods is $87.78\%$, which is a good agreement. Therefore, we conclude that for this specific network, where we have hashtags that are indicative of the side of the debate, both classification using hashtags and our classification method using sentiment and the network structure are possible and accurate, with our method still outperforming the hashtag classification.

\begin{table}[H]
\centering
\caption{Confusion matrix between our method of classification and the hashtag-based user classification. The balanced accuracy is $87.78\%$.}
\begin{tabular}{c|ccc}
\hline
 Membership / Hashtag & No & Yes\\
\hline
C1 (Yes) & 54 (4.00\%) & 1294 (96.00\%)\\
C2 (No) & 226 (79.58\%) & 58 (20.42\%)\\
\end{tabular}
\label{tab:cross_hash_memb}
\end{table}

However, for the Irish Marriage Referendum dataset, it is not possible to use the hashtag classification, since the hashtags contained in that dataset are both neutral --- not indicative of a side in the debate. As shown in Appendix~\ref{chap:IMR}, our method of classification performs well on that dataset, giving a balanced accuracy of $82.9\%$. Therefore, we show that our method of classification can be used in any polarised social media discussion, and does not rely on hashtag classification.

\section{The method using the followership network} \label{chap:follower}

The follower list of users was used to create a network combining information from mentions and followership, as done by O'Sullivan et al.~\cite{o2017integrating}, and then compare the results with the ones obtained by using only the mentions network. It was also used to infer parents and children among the different users who tweeted the same text by following Goel et al.'s~\cite{goel2016structural} information diffusion analysis method. We retrieved all users that compose the Repeal the $8^\textrm{th}$ dataset and, using the Twitter Academic Research API again, we gathered all users followed by each user. We retrieved the followers of $181\,225$ ($95.9\%$) out of the $188\,928$ users in the mentions dataset. The number of links in the followership dataset is over $18$ million. The difference between number of users gathered for the followership dataset in comparison with the mentions dataset is due to the method to retrieve followers from a user. It works only for present active accounts, i.e., if an user in the mentions dataset has deleted or deactivated their account, their followers cannot be retrieved. We also observe that $95.4\%$ of the users have a list consisting of 1 to 500 followers after filtering by followers that used at least one of the tracked hashtags. The user with the largest number of followers using the tracked hashtags have $12\,465$ of them.

\subsection{Community detection on the combined mentions + followership network}\label{chap:follower_comm}

As done by O'Sullivan et al.~\cite{o2017integrating}, we combine the mentions and the followers networks in order to detect ideological communities, therefore we can use the results as a comparison for how our new method using only the mentions network performs. We start by applying different community detection methods to the mentions network and to the followers network separately. Once again, the Leiden method~\cite{traag2019leiden} achieved the highest modularity values --- slightly higher than Louvain~\cite{blondel2008fast} --- but when carrying out the analysis to distinguish polarised communities, Louvain outperforms, therefore we chose to apply Louvain (and the reliability on the comparison between methods is higher). The modularity values when using Louvain are $0.270$ for the followers network and $0.288$ for the mentions network. We then filtered the users by the ones belonging simultaneously to the first five largest communities in the followers network and the first nine largest communities in the mentions network (only the communities with $20$ nodes or more).

We retrieve $39$ combined communities by analysing the interactions among the communities in the mentions network and the communities in the followers network. We then apply k-means to cluster similar communities together (as described in Appendix~\ref{chap:IMR}) and uncover four community clusters, being three Yes supporters and one No supporter. By comparing the users classification with the annotated data we used in Section~\ref{chap:communities}, our final balanced accuracy for this method using a combination of the followers and the mentions networks is $86.95\%$, lower than the $90.9\%$ achieved by using only the mentions network. This discrepancy may be due to deletions of accounts (which impacts the follower network), especially accounts of users that supported the No vote, where we observed the lower accuracy ($53.69\%$), and due to the extra users filtering when combining both networks, which impacts the capacity of the community detection method to retrieve a reasonable community structure.

\subsection{Cascades inferred by using the followership network}\label{chap:follower_cascades}

\begin{figure}[H]
     \centering
    \includegraphics[width=0.9\textwidth]{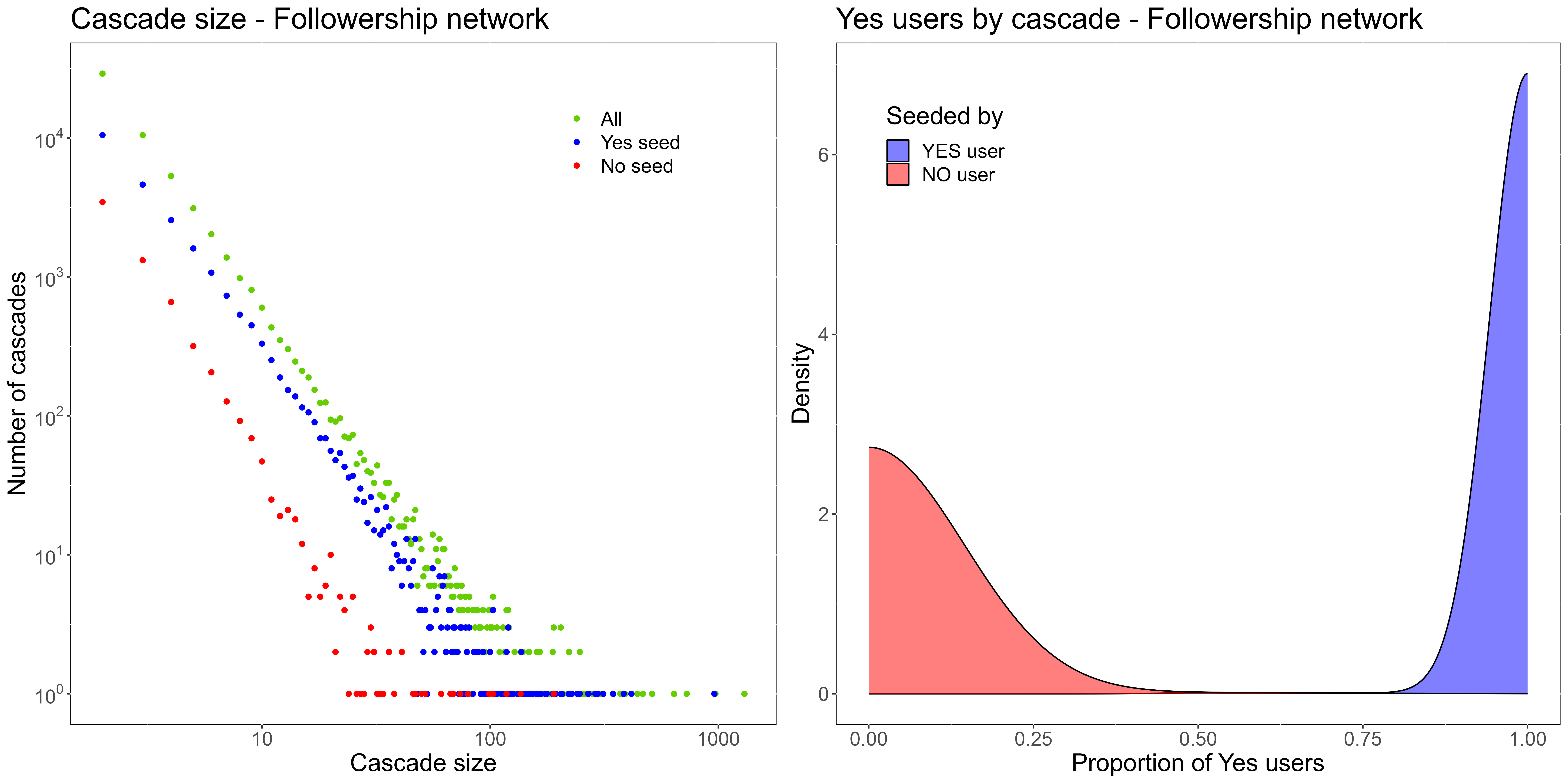}
    \caption{(a) Distribution of cascades sizes by seed community. (b) Density of Yes users by cascade tree, coloured by cascade seed type.}
        \label{fig:follower_diffusion}
\end{figure}

Following Goel et al.'s~\cite{goel2016structural} method of finding parenting, we use the followership network to check from whom a user is most likely to have retweeted from among the people they follow. By using this method, we retrieve $57\,430$ cascades from the $15\,340$ Retweet-IDs containing at least two tweets each. This shows that this method of parenting attribution does not perform as well as the method when using the mentions network (described in Section~\ref{chap:cascades}) --- where we were able to split the Retweet-IDs into only $24\,983$ cascades--- meaning that the process did not find suitable parents for a reasonable amount of retweets and considered a new cascade formation, even though the text was exactly the same. The results on the seeding analysis (Figure~\ref{fig:follower_diffusion}) are comparable to the ones achieved by using the mentions network as the network of inference. There is a higher number of small cascades in comparison to the results shown in Figure~\ref{fig:seed_sizes}~(a), and this is due to the fact that we split the Retweet-IDs into a larger number of smaller cascades when using the follower network for cascade inference. Figure~\ref{fig:follower_diffusion}~(b), however, shows the same qualitative results as those shown in Figure~\ref{fig:prop_yes_cascade}~(c). This clarifies that independent of which network we use to infer cascades, the polarisation is clearly affecting the way information spreads.

\section{The Irish Marriage Referendum} \label{chap:IMR}

The Same-sex marriage referendum in Ireland was held on 22 May 2015, and put Ireland in the history as the first country to approve same-sex marriage by popular vote~\cite{irishtimes2015Samesex}. The following text was inserted in the Irish Constitution:

\vspace{1cm}

\hfill\begin{minipage}{\dimexpr\textwidth-2cm}
\say{Marriage may be contracted in accordance with law by two persons without distinction as to their sex.}~\cite{murphy2016marriage}
\end{minipage}

\vspace{1cm}

The Yes vote prevailed by $62\%$ to $38\%$ with a large $60.5\%$ turnout in the referendum. In total, $1\,201\,607$ people voted in favour with $734\,300$ against, giving a majority of $467\,307$ votes. Only one of the Irish counties rejected the same-sex marriage, and the Yes vote was particularly pronounced in Dublin~\cite{irishtimes2015Samesex}.

The Irish Marriage Referendum data analysed in this work were collected by Sinnia, a data analytics company, using Twitter Gnip Power-Track API which returns a complete dataset, not just a sample~\cite{goel2016structural} as do other API, such as Twitter stream API~\cite{morstatter2013sample}. Every tweet containing the hashtags $\#marref$ and $\#marriageref$ from 8 May to 23 May 2015 (the day after the referendum) were collected.

\begin{figure}[H]
     \centering
         \centering
         \includegraphics[width=\textwidth]{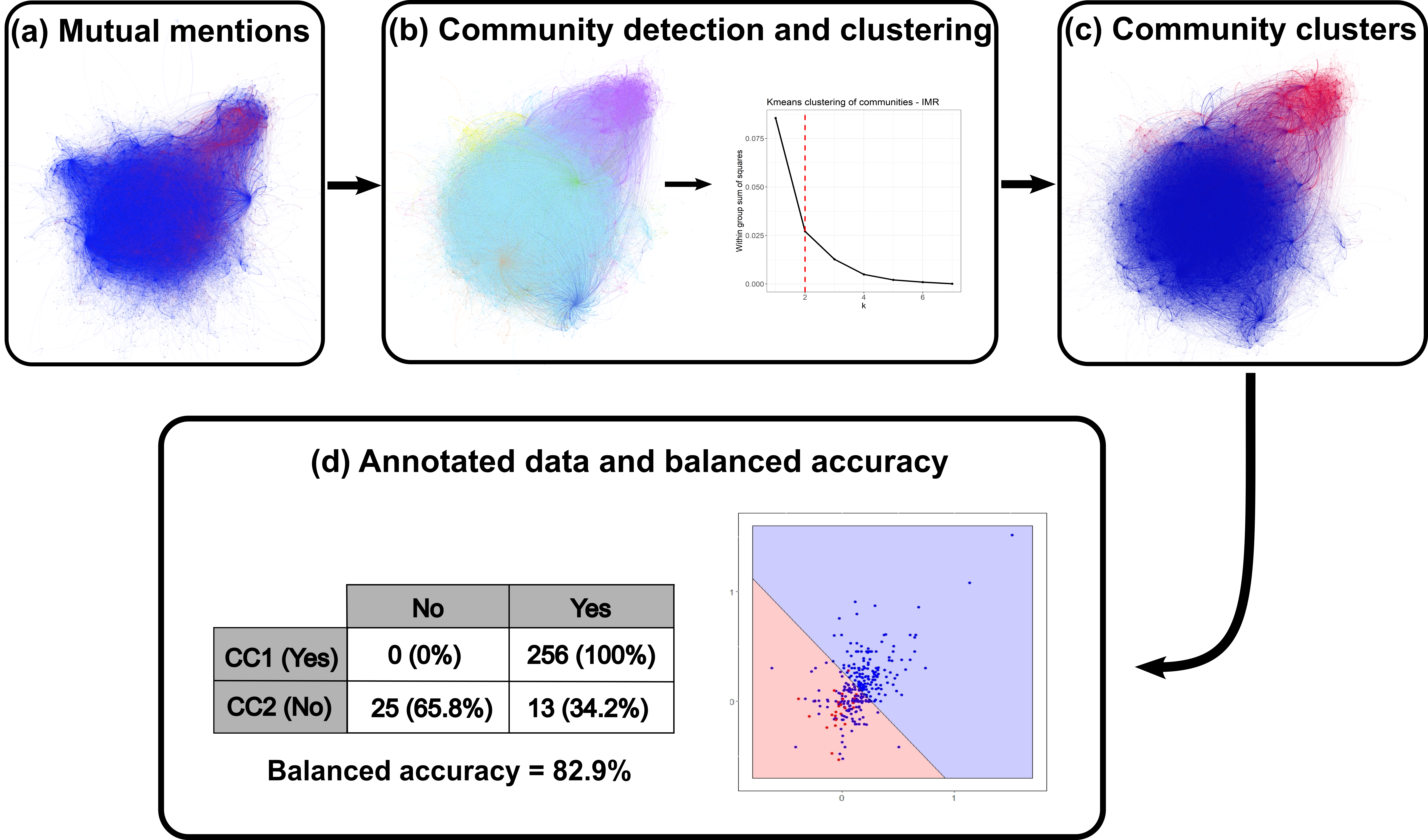}
        \caption{The community clustering process for the Irish Marriage Referendum mutual mentions network. (a) The mentions network before applying the community detection method (weighted Louvain), nodes and links coloured by sentiment (positive or negative); (b) Network visualisation with nodes and links coloured by the initial eight community memberships resulting from the community detection method; K-means clustering of detected communities where the parameters are the average sentiment-in and sentiment-out of each community. The optimal number of clusters is two; (c) How the initial eight communities are clustered into two final Community Clusters, which we nominate ``Yes community cluster'' and ``No Community Cluster''; (d) When analysing annotated data for side of the debate support, the majority of the Yes supporters lie on top of the Yes Community Cluster, and the opposite is also true. The balanced accuracy of our method of community identification is $82.9\%$.}
        \label{fig:IMR_network}
\end{figure}

As we did for the Repeal the $8^\textrm{th}$ analysis, we construct the mutual mentions network with averaged sentiment score by user. We use that network to find polarised communities by using the weighed Louvain~\cite{blondel2008fast} algorithm, for which the modularity is $0.23$. There are eight large (containing more than 20 nodes each) communities found by using the detection method, therefore we use k-means statistical clustering~\cite{lloyd1982least} to find a smaller number of final communities that could explain the network polarisation. By applying k-means we cluster the eight communities together into two community clusters (CCs) (Figure~\ref{fig:IMR_network}~(b)). The two final community clusters are well connected on the inside and less connected between each other (Figure~\ref{fig:IMR_network}~(c)), and when compared to the annotated data (manually classified users), we get good agreement (Figure~\ref{fig:IMR_network}~(d)), with a balanced accuracy of $82.9\%$, which is bigger than the balanced accuracy --- $81.0\%$ --- achieved previously~\cite{o2017integrating} when using a combination of the mentions and the followership networks.

\begin{figure}[H]
    \centering
    \includegraphics[width=0.9\textwidth]{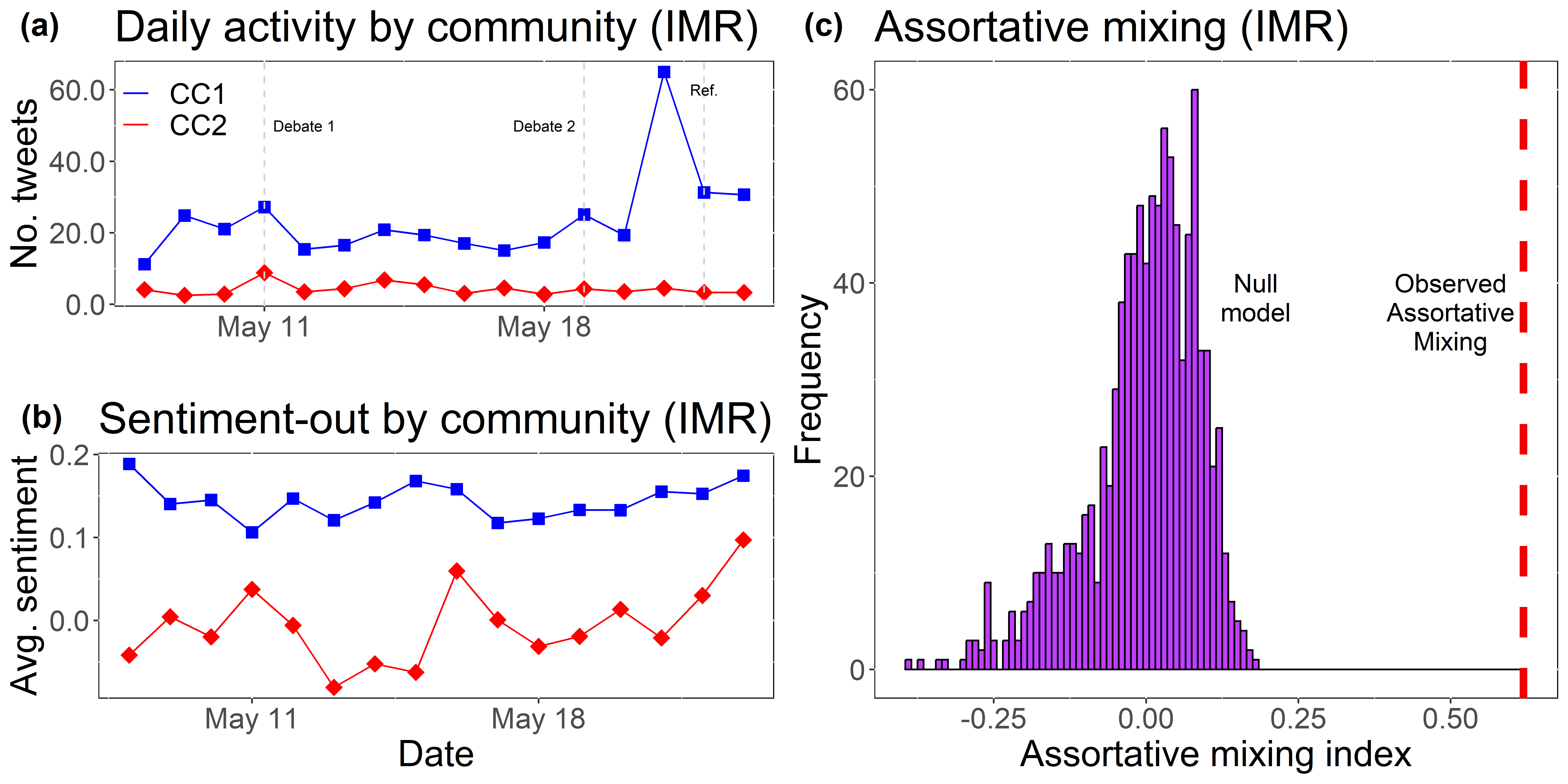}
    \caption{(a) Activity of the community clusters in the Irish Marriage Referendum discussion. The No Community Cluster has a peak (still weaker than the Yes Community Cluster) around the first televised debate, while the Yes CC has its peak of activity and is much more active than the No CC days before the Referendum day, when the majority voted Yes. (b) Sentiment-out of the community clusters in the Irish Marriage Referendum. Corroborating with the nomenclature we gave to the community clusters, the Yes CC (CC1) is constantly positive on average over time, while the No CC (CC2) is mostly negative or neutral on average over the time analysed, only leaning to a positive language after the referendum. (c) Monte Carlo simulation (histogram) of the Assortative Mixing index together with the observed Assortative Mixing index. The purple bars are the resultant Assortative Mixing index distribution obtained after 1000 simulations, and the red dashed line represents the observed index of $0.62$.}
    \label{fig:summary_CC_IMR}
\end{figure}

Following the polarised graph figure and the figure with the annotated data, we classify CC1 (blue) as the Yes community cluster, and CC2 (red) as the No community cluster. Figure~\ref{fig:summary_CC_IMR}~(b) corroborates with our classification by showing that CC1 presents a consistent positive average sentiment-out over time, and CC2 has a lower average sentiment-out over time. Figure~\ref{fig:summary_CC_IMR}~(a) shows that the activity (average number of tweets per user) is higher for CC1, and after debate 2 --- days preceding the voting --- the Yes community has a high peak, that is, Yes supporters tweeted a lot more, on average, than the No supporters on the days preceding the referendum. The observed assortative mixing index is greater and far away from the simulated ones (Figure~\ref{fig:summary_CC_IMR}~(c)), indicating that the assortative mixing does not occur due to chance alone, instead it depends on the network structure. The observed correlation of $0.62$ suggests a moderate to high assortative mixing uncovered by the community detection method.

We conclude that by using a simpler method of network creation and community detection when compared to O'Sullivan et al.'s~\cite{o2017integrating} method, we get successful results with a higher balanced accuracy, indicating that our simpler method, which uses less data, works better for the Irish Marriage Referendum data, as well as for the Repeal the $8^\textrm{th}$ data, as discussed in previous sections.

\section{Cascade scores} \label{chap:appendix_cascades}

A well-explored method to access how viral, or how diffuse a cascade tree ($T$) is, is the assignment of scores to it. Here we follow the discussion in Goel et al.~\cite{goel2016structural} to measure the success of a cascade. We will discuss the results obtained by using maximum cascade depth (Eq.~\ref{eq:max_depth}), average cascade depth (Eq.~\ref{eq:avg_depth}), and structural virality (Eq.~\ref{eq:virality}). The maximum cascade depth, the most intuitive method for assessing the ``success'' of a cascade, is not able to distinguish between a viral behaviour and a large broadcast with just one, long, multigenerational branch~\cite{goel2016structural}. The average cascade depth corrects for this issue. However, it fails if the information spreads through a long path from the root and then is broadcast out to a large group of adopters. In this case, the cascade would have high average depth (since most adopters are far from the root) even though most adoptions are the result of a single influential node. The structural virality, otherwise, is able to deal with both matters but it is possible to construct hypothetical examples for which it might fail, according to its authors~\cite{goel2016structural}.

\begin{equation}
    M(T) = \max\limits_{i} d_{ij},
    \label{eq:max_depth}
\end{equation}

where $d_{ij}$ denotes the shortest path between nodes $i$ and $j$ (the seed node) in a given cascade tree $T$.

\begin{equation}
    A(T) = \frac{1}{n}\sum_{i}^{n} d_{ij},
    \label{eq:avg_depth}
\end{equation}

where $j$ is the seed node and $n$ is the number of nodes in a given cascade tree $T$.

\begin{equation}
    V(T) = \frac{1}{n(n-1)}\sum_{i}^{n} \sum_{j}^{n} d_{ij}.
    \label{eq:virality}
\end{equation}

\begin{figure}[H] 
    \centering
    \includegraphics[width=0.8\textwidth]{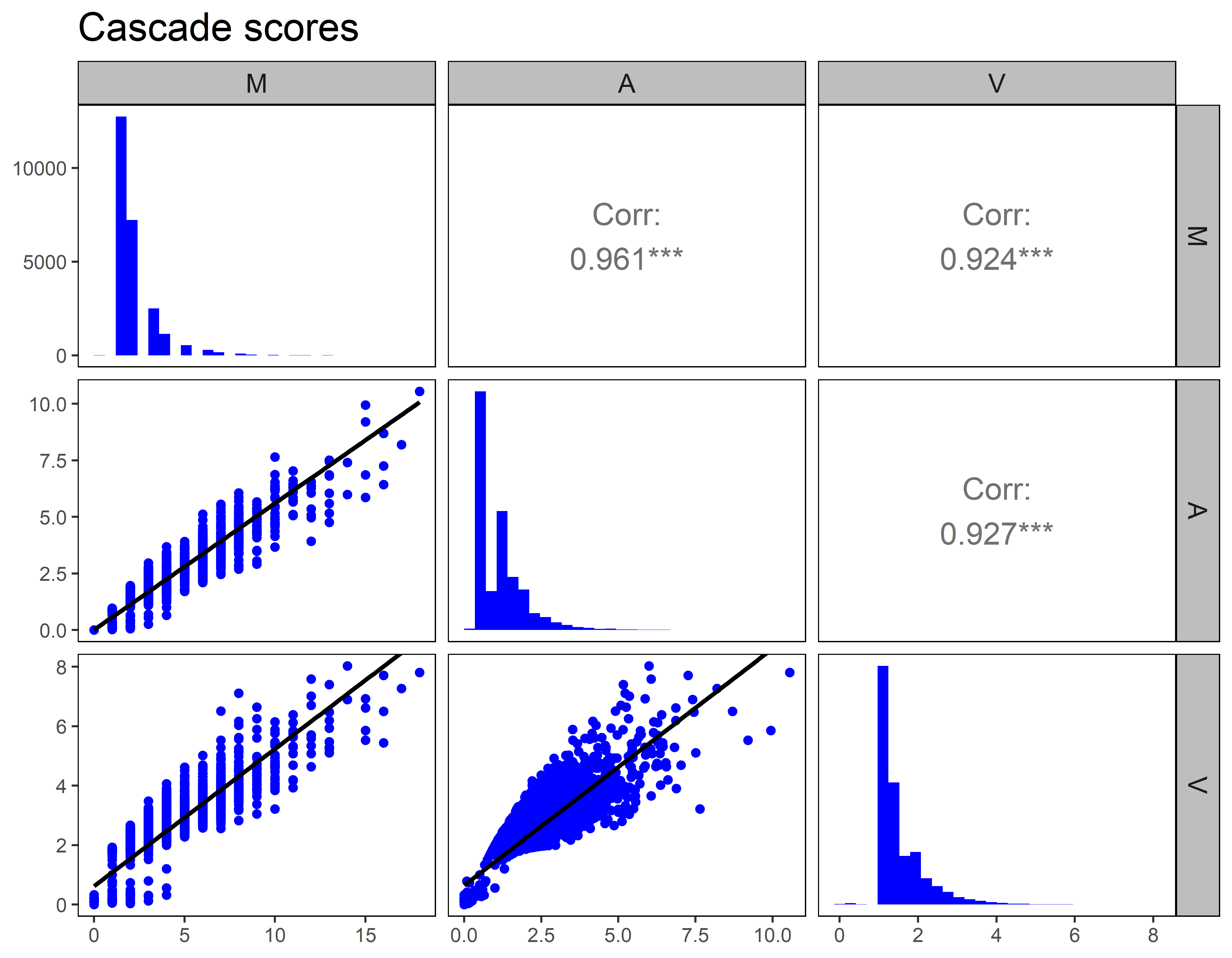}
    \caption{Pearson correlation, distribution plots and scatter plots of pairs of cascade scores. There is a statistically significant linear correlation between all measures presented above ($p<0.001$ in all three cases).}
    \label{fig:ggpairs_cascades}
\end{figure}

Figure~\ref{fig:ggpairs_cascades} shows the resultant distributions of scores of our retrieved cascades. There is high positive correlation between all pairs of scores, and the distributions have similar behaviour, i.e., all of them present a right (or positive) skew with mode $1.0$ for the maximum cascade depth, and $0.5$ for both average cascade depth and $1.0$ for structural virality, median of $1.0$, $0.8$ and $1.33$, and mean values of $1.89$, $1.06$ and $1.49$, respectively, meaning that the majority of the cascades present more of a broadcast behaviour (small scores). Figure~\ref{fig:seed_sizes} shows that the majority of cascades are small, with many presenting only one retweet (one user retweeted from the seed user). Those small cascades present low scores, bringing the median and the mean down.

\newpage

\bibliography{refs}

\end{document}